\documentclass[10pt,letterpaper]{article}
\usepackage[top=0.85in,left=2.75in,footskip=0.75in]{geometry}

\usepackage{amsmath,amssymb}

\usepackage{changepage}

\usepackage[utf8x]{inputenc}

\usepackage{textcomp,marvosym}

\usepackage{cite}

\usepackage{nameref,hyperref}

\usepackage{microtype}
\DisableLigatures[f]{encoding = *, family = * }

\usepackage[table]{xcolor}

\usepackage{array}

\usepackage{multirow}

\newcolumntype{+}{!{\vrule width 2pt}}

\newlength\savedwidth

\newcommand\thickhline{\noalign{\global\savedwidth\arrayrulewidth\global\arrayrulewidth 2pt}%
\hline
\noalign{\global\arrayrulewidth\savedwidth}}

\usepackage{setspace} 
\raggedright
\usepackage[aboveskip=1pt,labelfont=bf,labelsep=period,justification=raggedright,singlelinecheck=off]{caption}

\makeatletter
\renewcommand{\@biblabel}[1]{\quad#1.}
\makeatother

\usepackage{lastpage,fancyhdr,graphicx}
\usepackage{epstopdf}

\fancyheadoffset[L]{2.25in}
\fancyfootoffset[L]{2.25in}
\newcommand{\covid}{COVID-19}

\newcommand{\fref}[1]{Fig~\ref{#1}}

\newcommand{\tref}[1]{Table~\ref{#1}}
\newcommand{\eref}[1]{Eq~(\ref{#1})} 

\begin{document}
\vspace*{0.2in}

\begin{flushleft}
{\Large
\textbf\newline{A generalized forecasting solution to enable future insights of \covid\ at sub-national level resolutions} 
}
\newline
\\
Umar Marikkar\textsuperscript{* 1\Yinyang},
Harshana Weligampola\textsuperscript{1\Yinyang},
Rumali Perera\textsuperscript{2},
Jameel Hassan\textsuperscript{1},
Suren Sritharan\textsuperscript{3},
Gihan Jayatilaka\textsuperscript{1},
Roshan Godaliyadda\textsuperscript{1},
Vijitha Herath\textsuperscript{1},
Parakrama Ekanayake\textsuperscript{1},
Janaka Ekanayake\textsuperscript{1},
Anuruddhika Rathnayake\textsuperscript{4},
Samath Dharmaratne\textsuperscript{5},
\\
\bigskip
\textbf{1} Faculty of Engineering, University of Peradeniya, Sri Lanka
\\
\textbf{2} Faculty of Science, University of Peradeniya, Sri Lanka
\\
\textbf{3} Sri Lanka Technological Campus, Padukka, Sri Lanka
\\
\textbf{4} Postgraduate Institute of Medicine, University of Colombo, Sri Lanka
\\
\textbf{5} Faculty of Medicine, University of Peradeniya, Sri Lanka
\\
\bigskip

%
%
\Yinyang These authors contributed equally to this work.





*Corresponding author: umar.m@eng.pdn.ac.lk

\end{flushleft}
\section*{Abstract}

\covid\ continues to cause a significant impact on public health. To minimize this impact, policy makers undertake containment measures that however, when carried out disproportionately to the actual threat, as a result if errorneous threat assessment, cause undesirable long-term socio-economic complications. In addition, macro-level or national level decision making fails to consider the localized sensitivities in small regions. Hence, the need arises for region-wise threat assessments that provide insights on the behaviour of \covid\ through time, enabled through accurate forecasts. In this study, a forecasting solution is proposed, to predict daily new cases of \covid\ in regions small enough where containment measures could be locally implemented, by targeting three main shortcomings that exist in literature; the unreliability of existing data caused by inconsistent testing patterns in smaller regions, weak deploy-ability of forecasting models towards predicting cases in previously unseen regions, and model training biases caused by the imbalanced nature of data in \covid\ epi-curves. Hence, the contributions of this study are three-fold; an optimized smoothing technique to smoothen less deterministic epi-curves based on epidemiological dynamics of that region, a Long-Short-Term-Memory (LSTM) based forecasting model trained using data from select regions to create a representative and diverse training set that maximizes deploy-ability in regions with lack of historical data, and an adaptive loss function whilst training to mitigate the data imbalances seen in epi-curves. The proposed smoothing technique, the generalized training strategy and the adaptive loss function largely increased the overall accuracy of the forecast. The fact that forecasts made on regions with the generalized training strategy actually outperformed the current practice of using local training data is an important finding of this work. Therefore, despite the absence or limited existence of local pandemic data, the proposed methodology enables efficient containment measures at a more localized micro-level.



\section*{Introduction}

The \covid\ pandemic has now spread to all regions throughout the world. Classified as a global pandemic by the World Health Organisation (WHO), \covid\ continues to affect humankind more than a year after its first recorded case, causing over 4.3 million total recorded deaths worldwide as of August 2021, with close to ten thousand deaths per day \cite{JHU_dashboard}. This gradual increase in numbers has pushed governments and policy makers to enforce containment measures that include restrictions on public gatherings, local and international travel bans, and region-wise lock-downs\cite{travel_restrictions,lockdown}. 
Although these policies have attenuated the effect of \covid\ from a mere numbers standpoint, they have given rise to a variety of side-effects from multiple viewpoints\cite{pone_mental_lockdown, eci_non-evidence, pone_heartrate_lockdown}. Studies have shown the existence of a significant psychological impact on students due to lack of social interactions caused by distance-learning programs carried out by schools and universities around the world, such as USA\cite{pone_mental_NJ,pone_psychological_US}, Bangladesh\cite{pone_depression_BD} and Spain\cite{pone_edu_spain}. From an economical perspective, these measures have led to closure of industries and small businesses, negatively affecting those who rely on their daily or weekly income for self-sustenance\cite{small_businesses}. 
Furthermore, the oscillatory behaviour of the \covid\ pandemic caused by what is commonly termed as covid "waves" calls for policy makers to adapt and optimize containment measures in response to current severity levels. This is because a disproportionate response while attenuating disease spread will unnecessarily create an adverse socio-economic impact, hence the need for proportionate responses based on threat levels. In addition to this oscillatory behaviour, the localized nature of \covid\ spread also calls for responses to be enacted locally, in a way that is unique to each sub-region as opposed to generalizing over a larger region. \par

This need for proportionate responses can be fulfilled upon optimal decision making on containment policies enforced by policy makers, achieved via data-driven analysis of \covid\cite{optimal_decisions, dzau2020strategy}, that attempt to optimally balance disease transmission mitigation with the aforementioned socio-economic impact cost. Different governments use different policies when the number of cases rises or an outbreak occurs within the country. For example, a group of experts have proposed a national framework for the USA, which contains four sets of policies depending on the number of new confirmed \covid\ cases per one million population on a single day \cite{alert_USA}. Similarly, the Centre for Disease Prevention (CDC) of the USA has defined risk levels of \covid\ to determine travel restrictions across regions,  to optimally balance of the transmission mitigation versus socio-economic impact trade-off\cite{alert_CDC}. These discrete threat level mechanics (or alert levels) are introduced as an effort to implement strategies to mitigate \covid\ transmission based on severity of the pandemic on a given location (country/state/province) as it allows the government and policy makers to initiate a proportionate response to curb the spread. Hence, ensuring that the adverse socio-economic impact is also optimally mitigated.  \par

For these strategies to be carried out optimally, there needs to be an understanding of how the pandemic spread will behave in the future, as it will be possible to analyze the long-term impact caused by applying certain policies based on existing threat levels. This understanding is reached by localized micro-level forecasts of \covid\ which helps obtain future insights on a qualitative basis, where numerical data is interpreted as actionable information, thereby enabling governments to initiate protocols in the aforementioned optimized manner.\par

Mathematical modelling of \covid\ through time serves as the basis for \covid\ forecasting studies. It is divided into two main sections; epidemiological models such as Susceptible-Infected-Recovered (SIR) models\cite{weiss2013sir} and their variants, and numerical forecasts obtained using AI (Artificial Intelligence) techniques such as Neural Networks (NN) and Deep Learning (DL) techniques \cite{zeroual2020deep}. Existing epidemiological models provide a broader, more generalized idea about the future behaviour of the disease. In contrast, the latter uses immediate historical data of epi-curves to obtain shorter, but more accurate numerical forecasts, thereby providing insights of infection data on a higher resolution, in turn playing a key role in assessing the state of the pandemic. Hence, Let us explore the level of work thus far carried out in numerical forecasts of \covid. \par

Many researchers have reviewed the use AI models in \covid\ forecasting \cite{AI_forecasting_review, jayatilaka2020use, IEEE_TAI}. These studies describe the use of statistical models such as AR and ARIMA, along with NN/DL models like Artificial Neural Networks (ANNs) and Long-Short-Term-Memory (LSTM) networks. It was observed here that the LSTM network was most commonly used, and at most times the best performing model for numerical forecasting of \covid. Upon region-wise analysis, the aforementioned AI techniques have been used to predict future \covid\ cases in many regions such as Canada \cite{pone_canada}, Pakistan \cite{pone_pakistan}, Brazil\cite{forecast_brazil} and many others\cite{forecast_EU,forecast_india}. These studies have been carried out for highly populated areas, and the forecasting was performed at country level where the epi-curves are smooth and data is abundant. However, upon searching for data and forecasting done at higher resolutions (i.e., at state/province/district/county level) the data and studies are lacking. It is evident that there is a clear need for region-wise forecasting at a higher resolution as the threat level for different regions in a single country might vary. Hence, in order to practice the optimum containment strategies argued previously, each state or province might require its own unique forecast and corresponding containment strategy. As, a nationwide threat level might be misleading and lead to an over-reaction or under-reaction by the government through its containment efforts, thus, compromising the entire effort.  For instance, although research has been done using statistical models to predict the total number of cases in Sri Lanka \cite{pone_SIR_srilanka}, there exists no district-wise prediction study within the country, which would be useful for policy makers to make more localized decisions in terms of medical resource allocation (hospitals, ICU beds and PPEs) and containment (through social distancing and motion control regulations and vaccination efforts). These localized prediction studies will be of paramount importance especially for countries with limited or strained resources as a result of the more contagious variants that have surfaced in recent times. \par

The scarcity in higher resolution localized forecasting studies is mainly due to the lack of reliable, deterministic historical \covid\ data that can be extracted into useful information, thereby increasing the difficulty to create rich datasets to be fed into AI models. A significant contributor towards this lack of reliable data is low testing frequencies and inconsistent testing patterns in some regions, resulting in unrealistic epi-curves where large fluctuations are displayed within relatively small periods\cite{failed_forecasting}. In addition to the lack of reliability in forecasting, a threat level assessment performed here too will be highly unreliable with raw data due to the same reasons. \par
To overcome the aforementioned issue of large fluctuations, multiple smoothing methods exist in literature, where noisy signals (i.e., fluctuating signals) are smoothed along the time domain. One such method, although relatively primitive, but used in the case of \covid, is the N-day averaging algorithm, where the current number of cases is the average value of the number of cases in N previous days \cite{JHU_dashboard}. An improved version of N-day averaging is the Moving Average (MA) filter, where the current number of cases is a weighted average of the cases in the past \cite{MA_filters}. The MA filter learns the dependency of the previous days' cases to determine the current number of cases. The use of smoothing algorithms in \covid\ forecasting is however limited to the two aforementioned methods; N-day averaging and MA filters, as alternative smoothing methods beyond these rudimentary techniques are yet to be explored in data pre-processing for the \covid domain. One such method, derived from Digital Signal Processing (DSP) literature, is low-pass filtering (LFP), led by the analysis of the original signal in its frequency domain. LPF selectively removes fluctuating components from a time-series signal, resulting in a much smoother signal. This is enabled by the conversion of the time-series signal from the time domain to the frequency domain, where the fluctuating nature of the signal can be directly quantified. The use of LPF is advantageous as it preserves the quality of information in the low-frequency (smooth) components of a signal, whilst selectively filtering its high-frequency (fluctuating) components\cite{LPF}. \par
However, LPF, N-day averaging, or MA, when used for \covid\ prediction and analysis displays a limitation, especially when applied for case studies that contain regions that exhibit a wide range of epi-curve patterns; some extremely fluctuating and some less so. The diverse nature of epi-curves results in sub-optimal smoothing when global smoothing parameters are used; where these parameters of a given algorithm are constant throughout the case study. The local context of epi-curves in a case study might add more or less localized/situational volatility, hence, a global smoothing might actually lead to information loss in such cases, as opposed to information enhancement that is needed to be achieved. For example, 3-day averaging may not be sufficient to smoothen epi-curves in a region where testing is carried out every 7 days, whereas 14-day averaging may cause information loss. Also, the number of tests carried out in each day will not be consistent. To address these problems, it is possible to manually set the smoothing parameters for each epi-curve in a case study. However, it could be a tedious task when given multiple datasets with a large number of epi-curves, and manual selection of these parameters might be subjected to operator bias. This gives rise to the requirement of an automated algorithm, where the optimal parameters of the smoothing technique are computed for each epi-curve, depending on their fluctuating nature. 

\par In this paper, an optimized LPF-based smoothing technique is proposed, which is tuned such that it attempts to smoothen a given epi-curve by maximizing the removal of fluctuating components, whilst minimizing the loss of useful information in the original signal in a local context. Upon the initial comparison of the proposed optimized smoothing technique with non-optimized techniques, thus verifying the need for automated, optimized techniques over manual selection of smoothing parameters. Then, the feasibility of the proposed localized smoothing technique was validated by comparing \covid\ alert levels in a given region using the proposed smoothing method versus unsmoothed epi-curves. The alert level analysis was carried out based on a known alert level system\cite{alert_USA}, where raw and smoothed case data was converted to \covid\ alert levels, and the effect on reliability of the alert levels in that region due to the proposed smoothing technique was discussed. To compute the alert levels, two distinct methods were used, one which responds instantly to highly fluctuating data, and the other which hardly responds to fluctuating data. The contrasting characteristics of these alert levels were used to highlight the importance of the optimized smoothing technique. Upon validation through alert level analysis, the proposed technique was used to smoothen epi-curves which were fed as training data to train a \covid\ daily case forecasting model. This would then firmly establish the adaptive localized smoothing operation as the optimum data conditioning tool to be used under high resolution local environments prior to performing forecasting tasks. Upon carrying out the forecasting process, it was also found that the forecasting models trained using \covid\ data smooothed using the proposed smoothing algorithm resulted in significantly higher prediction accuracies, even upon evaluation with raw data as the ground truth.\par
Another key drawback observed in existing literature for forecasting \covid\ \cite{pone_canada, pone_pakistan, forecast_brazil, forecast_EU, forecast_india} is that most commonly, data from a particular region is used for both training and testing purposes, and their corresponding forecasting models. This leads to a decrease in the deploy-ability of these models, by limiting the possibility of these models being used to forecast \covid\ cases in regions that are un-trained by them. For instance, a model trained using \covid\ cases in India will not perform well in predicting \covid\ cases in a region such as the USA, due to the contrasting nature of their epi-curves \cite{cooper2020sir}. Especially with the advent of new variants, countries previously less affected by the pandemic are observing sudden surges in infections and deaths. However, the past infection data on many of these countries would hardly provide any insight onto how the epidemic can be forecasted in its current state. This is amplified by the fact that certain countries also only perform testing in limited capacities. Hence, the need arises for forecasting solutions trained in a more generalized/global manner as these can now be used for such nations. This will become even more prominent with more and more variants coming into play. \par

Such a solution is proposed in this study, through a generalized training strategy, where a \covid\ forecasting model is trained using a select number of `diverse' and `representative' regions, such that the model will have the ability to predict \covid\ cases in any given region. The choice of these training regions was subject to the diversity in geographical location and demographic features of each region, as these representative regions on which the forecasting model is trained have to form a diverse enough basis both geographically and demographically to train the model such that it is usable anywhere in the world. As for the forecasting model, Long-Short-Term-Memory (LSTM) based Neural Network (NN) was designed. The use of NNs for time-series forecasting for \covid \cite{AI_forecasting_review} coupled with the abundant use of LSTMs for time-series problems \cite{DL_time_review} led to an LSTM-based NN being the optimal forecasting architecture to be used for our case study. This predictive model was evaluated by forecasting daily new cases in test regions previously unseen by the model (i.e., not used in training). \par

A significant finding of this study is that choosing the aforementioned generalized training strategy to train the LSTM-based NN model results in better prediction performances when compared to models trained using previously observed data from the test region itself. Therefore, it is recommended to train the LSTM-based NN model through a geographically and demographically representative set of countries to form a basis that would ideally span the total possibility space. Then when a new region is selected for forecasting through a model trained in this manner, it would account for most the possibilities the new wave the region in concern would encounter. Therefore, if it sees that specific attribute in the region’s epidemiology model it would trigger a prediction based on that aspect. Hence, resulting at times in better prediction performances that compared to models trained using previously observed data from the region itself.\par

Poorly trained NN models result in poor and inaccurate predictions. A factor that negatively affects the performance of NN models in the training process is the lack of balanced data in a training dataset. In the instance of \covid, considering daily new cases throughout a long period of time for a region/country, the imbalance is caused by most values in an epi-curve being zero. This is due to either \covid\ cases being recorded periodically (resulting in the non-recorded days being zero), or an instance where an epi-curve has a multi-modal nature (due to separate waves of the pandemic), resulting in no recorded cases between two peaks. This large number of zero values results in a training bias, where the NN model would be more inclined to predict zeros due to having 'seen' more zeros in the training process. To overcome this, a density-based adaptive loss function \cite{undersampling_loss_combine} was proposed. The performance the final LSTM-based NN model consisting of the adaptive loss function was evaluated, and compared with models trained with the standard loss functions. It was observed that the model trained using the proposed loss function produced more accurate predictions when compared to the models trained using standard loss functions, specifically for the models trained with raw (unsmoothed) data.\par

In summary, the contributions of this paper are as follows, 
\begin{itemize}
    \item An optimized smoothing technique to smooth \covid\ epi-curves in regions with varying testing patterns was proposed using time-domain and frequency-domain analysis of signals. The smoothing technique is locally/contextually optimized based on the epidemiological dynamics of the region allowing for better data condition prior to use for forecasting and modeling. 
    \item  An analysis of behavioural patterns of different alert level systems was carried out for smoothed vs. non-smoothed epi-curves.
    \item An LSTM-based NN predictive model to predict daily new \covid\ cases 10-days ahead, trained using a generalized training strategy to construct a more representative training space/basis was introduced that is capable of predicting \covid\ infection levels for any given region. This includes regions not included in the training set, hence, allowing for predictions in regions with lack of data and/or with new waves of infections. 
    \item An adaptive loss function was proposed for the LSTM-NN based predictive model to mitigate the common problem of high zero values in epi-curves. 
\end{itemize}

\section*{Materials and methods}

\begin{figure}[!ht]
    \centering
    \includegraphics[width=\textwidth]{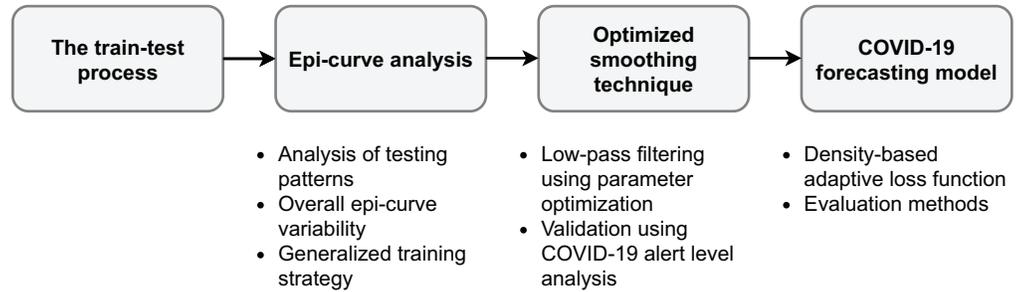}
    \caption{\bf Summary of methodology}
    \label{fig:schematic_method}
\end{figure}

The Materials and Methods section is summarized as follows. First, an introduction to the train-test process is given. Then, the proposed optimal smoothing technique is presented, followed by a description of the \covid\ forecasting model. Finally, forecasting model was evaluated using the proposed generalized training strategy. A schematic flow of the methodology is shown in \fref{fig:schematic_method}.\par

\subsection*{The train-test process}

\begin{figure}[!ht]
    \centering
    \includegraphics[width=\textwidth]{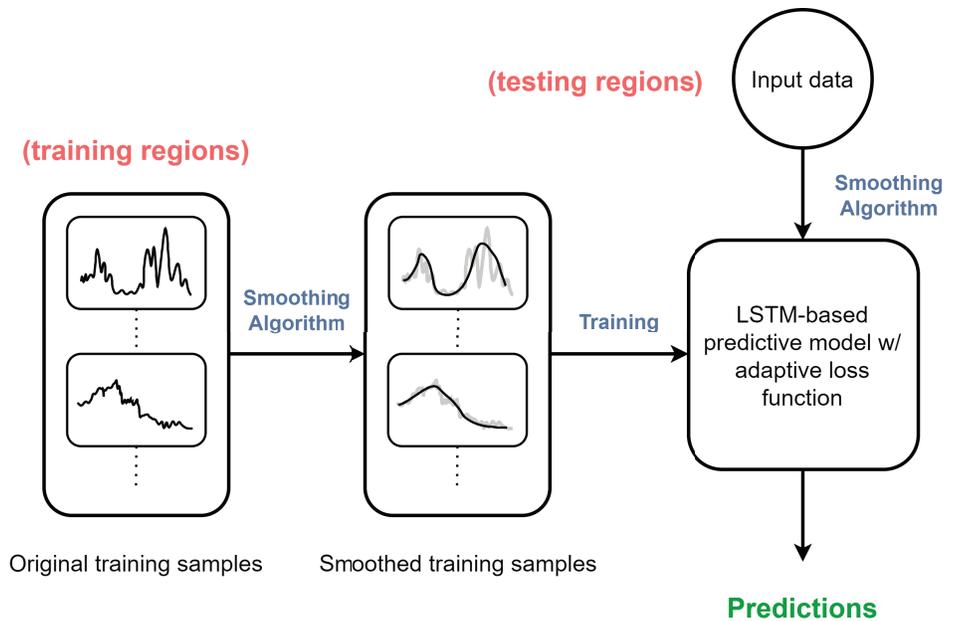}
    \caption{\bf The train-test process}
    \label{fig:train_test_process}
\end{figure}

Our work focuses on developing an end-to-end generalized solution towards forecasting daily new cases of \covid\ in any given region throughout the globe, on a localize, high-resolution context. In simple terms, a generalized approach towards localized forecasts of \covid. This solution consists of a training and testing process. During the training process, a universally optimized smoothing function is used to smoothen noisy epi-curves, followed by a training of the time-series forecasting model for \covid\ daily cases that employs a generalized training strategy to train an LSTM-based neural network along with the density-based adaptive loss function. The testing process evaluates the trained LSTM-based predictive model on previously unseen datasets, where historical \covid\ data is of the regions belonging to these datasets is not considered during any part of the training process. The train-test process is further described in \fref{fig:train_test_process} through a simple schematic diagram. \par 

\subsection*{Epi-curve analysis and datasets used}

The datasets in our case study were divided into two sections; training and testing. Training data would be used in setting the model parameters in the LSTM-based predictor and tuning the smoothing algorithm, whereas testing data was used to validate the forecasting process, thereby providing an insight as to how the model reacts to previously unseen data.\par

\begin{figure}[!ht]
    \includegraphics[width=\textwidth]{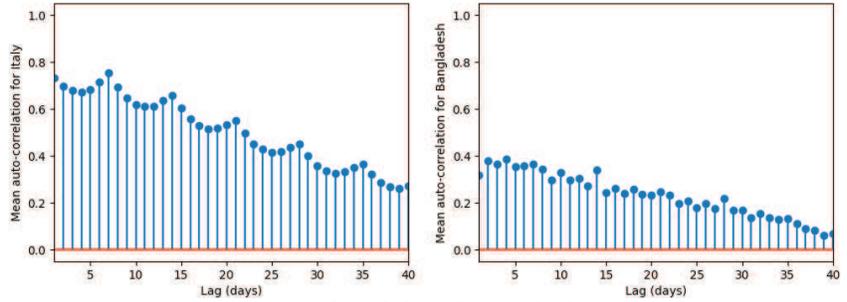}
    \caption{\bf Mean auto-correlation for COVID-19 daily cases in sub-national regions.} (left) provinces in Italy. (right) districts of Bangladesh.
    \label{fig:autocorr}
\end{figure}

First, let us consider a few examples of epi-curve variabilities, which will establish the need for the generalized training strategy and the optimized smoothing algorithm. The presence of variations in epi-curve properties across regions was important to encapsulate most kinds of possible epi-curve scenarios in the training process. One such example would be the testing patterns in each region. Consider the auto-correlation functions of epi-curves of 2 regions from the training dataset; Italy and Bangladesh, shown in \fref{fig:autocorr}. An auto-correlation function determines how much of a current value in time is dependent on its previous values. In the case of epi-curves of provinces in Italy, a peak in auto-correlation is observed every 7 days. In other words, the number of daily cases on a given day of the week correlates better with the same day of the previous weeks, thus indicating a weekly testing arrangement being carried out for each province. However, epi-curves of districts in Bangladesh does not show such trends; a result of a less deterministic testing system within the country, an attribute consistent throughout countries of the South Asian region.\par

Another factor, apart from the testing frequencies shown in \fref{fig:autocorr}, that mainly solidifies the need for an optimal smoothing algorithm is the quality of testing patterns in geographical areas. This determines the amount of data that can be extracted as useful information, in terms of \covid\ cases or deaths. This is due to epi-curves containing inconsistent testing patterns result in higher fluctuations that contribute to its noisy nature. This implies that there exists a variation in the smoothness of epi-curves in each area.

\begin{figure}[!ht]
    \includegraphics[width=0.6\textwidth]{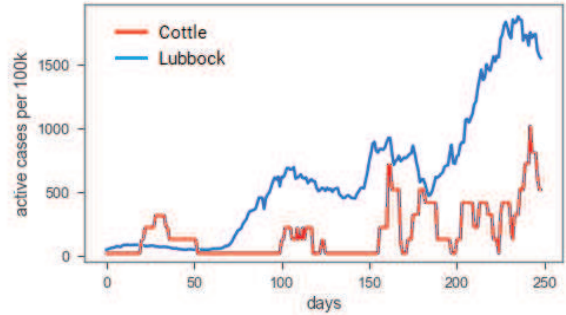}
    \caption{\bf Daily \covid\ cases for Cottle County and Lubbock County.}
    \label{fig:texas_testing}
\end{figure}

For example, consider two counties in the state of Texas, Cottle county (with a population of 1,642 in 2019) and Lubbock county (with a population of 310,569 in 2019). Since only total number of tests in each county is available (lack of testing data: a trait consistent with most smaller sub-regions), the relationship between the daily new \covid\ cases with number of tests carried out in each day cannot be derived. Therefore, it was assumed that there exists a correlation between the daily testing patterns and the total number of tests. This is inferred through visual observation in \fref{fig:texas_testing}, where Cottle County and Lubbock County, having 431 and 57761 total number of tests respectively (as of January 2021), differ in the fluctuating nature of their respective epi-curves, possibly due to the differences in daily testing patterns for each county. In addition to the aforementioned two sub-regions, the contrasting nature of epi-curves in the wide range of sub-regions used in this study is infinite, due to their geographic and demographic diversity, thereby further clarifying the need for optimal smoothing and the proposed generalized training strategy. \par

\subsubsection*{Training regions}

\begin{figure}[!ht]
     \includegraphics[width=0.7\textwidth]{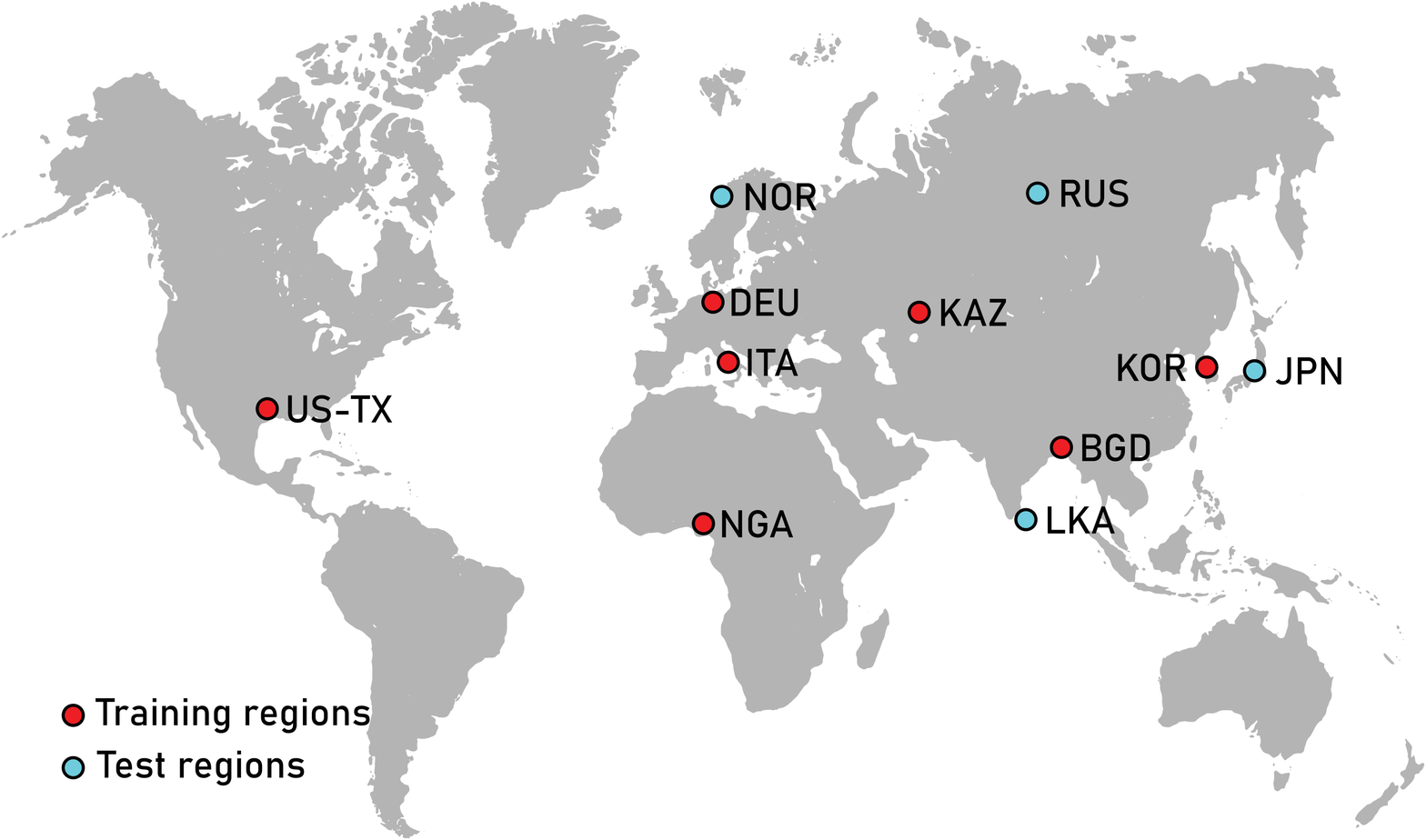}
    \caption{\bf The geographical spread of regions used for training and testing the proposed forecasting model}
    \label{fig:region_map}
\end{figure}

For training, sub-national \covid\ daily case data were obtained from a variety of regions, both demographically and geographically diverse from each other. This allows for a robust model that maximizes deployability for any region throughout the world. Data was collected from Counties in the state of Texas, USA (US-TX) and sub-national data was collected from states in Nigeria (NGA), provinces in Kazakhstan (KAZ), provinces in Italy (ITA), districts of Bangladesh (BGD), cities in Korea (KOR) and states of Germany (DEU).  To obtain more variations between epi-curves through geographic diversity, regions belonging to the training dataset were chosen such that these regions are spread out throughout the world, as shown in \fref{fig:region_map}.\par

\subsubsection*{Testing regions}

As opposed to using part of data from each country as test data, daily \covid\ cases reported from previously unseen regions were used to validate the model. Predicting future \covid\ cases in these 'unseen' regions will ensure the deploy-ability of the model whilst validating the proposed generalised training strategy.\par

In this case, sub-national \covid\ data from prefectures in Japan (JPN), cities in Russia (RUS), states of Norway (NOR) and districts of Sri Lanka (LKA) was used. It should be noted that the aforementioned regions were not at all used to train the LSTM-based NN model. This emulates a more realistic scenario where \covid\ predictions have to be carried out on a completely new region plagued with the pandemic, or a case where historical \covid\ data is lacking. On top of this, new variants such as the Delta and epsilon variants bring about new epi-curve dynamics. The transmissiblity of these variants contribute to the general rate of increase that can be seen in epi-curves. Therefore, this enables us to incorporate the surge dynamics of such highly contagious variants from regions first affected by it into regions that just got exposed to it, which enables rapid response for these regions. The geographical locations of each test region are also depicted in \fref{fig:region_map}. \par

To validate the smoothing algorithm, epi-curves from Sri Lanka and Texas were extracted, whilst Sri Lanka alone was used to analyze how \covid\ alert levels were obtained after smoothing. The reason for this is that Sri Lanka does not have an ongoing alert level system, hence it is of paramount importance that a usable alert level system for Sri Lanka is formulated. A summary of all datasets and their uses in each case study is summarized in \tref{tab:case_study}.\par


\begin{table}[!ht]
\begin{adjustwidth}{-2.25in}{0in} 
\centering
\caption{\bf Summary of datasets and their use in each case study}
\resizebox{\linewidth}{!}{

\begin{tabular}{|l|l|l|l|l|l|l|l|l|l|}
\hline
\multicolumn{2}{|l|}{\textbf{Region}} & \multicolumn{4}{l|}{\textbf{Features}} & \multicolumn{4}{l|}{\textbf{Case study}} \\ \hline
\multirow{2}{*}{\textbf{Name}} & \multirow{2}{*}{\textbf{ISO code}} & \multirow{2}{*}{\textbf{Resolution}} & \multirow{2}{*}{\textbf{\begin{tabular}[c]{@{}l@{}}No. of \\ epi-curves\end{tabular}}} & \multirow{2}{*}{\textbf{Start date}} & \multirow{2}{*}{\textbf{End date}} & \multirow{2}{*}{\textbf{Smoothing}} & \multirow{2}{*}{\textbf{\begin{tabular}[c]{@{}l@{}}Alert-level \\ analysis\end{tabular}}} & \multicolumn{2}{l|}{\textbf{Forecasting}} \\ \cline{9-10} 
 &  &  &  &  &  &  &  & \textbf{Train} & \textbf{Test} \\ \hline
\textbf{Texas} & \textbf{US-TX} & Counties & 254 & 04/04/2020 & 02/09/2021 & \checkmark &  & \checkmark &  \\ \hline
\textbf{Nigeria} & \textbf{NGA} & States & 37 & 02/27/2020 & 05/03/2021 &  &  & \checkmark &  \\ \hline
\textbf{Italy} & \textbf{ITA} & Provinces & 149 & 02/24/2020 & 05/24/2021 &  &  & \checkmark &  \\ \hline
\textbf{Bangladesh} & \textbf{BGD} & Districts & 64 & 07/08/2020 & 01/15/2020 &  &  & \checkmark &  \\ \hline
\textbf{Kazakhstan} & \textbf{KAZ} & Provinces & 17 & 03/27/2020 & 05/30/2021 &  &  & \checkmark &  \\ \hline
\textbf{Korea} & \textbf{KOR} & Cities & 17 & 01/20/2020 & 05/20/2021 &  &  & \checkmark &  \\ \hline
\textbf{Germany} & \textbf{DEU} & States & 16 & 01/03/2020 & 05/23/2021 &  &  & \checkmark &  \\ \hline
\textbf{Japan} & \textbf{JPN} & Prefectures & 47 & 03/18/2020 & 05/19/2021 &  &  &  & \checkmark \\ \hline
\textbf{Sri Lanka} & \textbf{LKA} & Districts & 26 & 11/14/2020 & 03/19/2021 & \checkmark & \checkmark &  & \checkmark \\ \hline
\textbf{Russia} & \textbf{RUS} & Cities & 83 & 03/20/2020 & 05/04/2021 &  &  &  & \checkmark \\ \hline
\textbf{Norway} & \textbf{NOR} & States & 11 & 01/03/2020 & 05/23/2021 &  &  &  & \checkmark \\ \hline
\end{tabular}
}
\label{tab:case_study}
\end{adjustwidth}
\end{table}

\subsection*{Optimized smoothing technique}
\label{sec:smoothing}

\subsubsection*{Low-pass filtering using parameter optimization}

Low pass filters are utilized to denoise noisy signals such as in Cottle country (as in \fref{fig:texas_testing}, by modifying the signal in its spectral (frequency) domain.  A low pass filter essentially attenuates these high-frequency components whilst keeping the low-frequency components intact. One of the main parameters of a low-pass filter is its cutoff frequency. It determines the degree to which a given signal will be smoothed/filtered. In the case of Texas, if all counties are filtered equally (i.e., with a constant cutoff frequency), there exists the possibility of under-filtering or over-filtering data. A smoothed signal, when under-filtered, contains a significant number of noisy components that remain from the original signal. In contrast, when a signal is over-filtered, it loses most of its useful information in the filtering process possibly resulting in an over-smoothed signal.\par

An optimized low-pass filter model is proposed, which attempts to minimize its noisy nature while maximizing the amount of useful information retained. This condition is determined by an optimal cutoff frequency at which the low-pass filter operates to denoise epi-curves in each region. A first order Butterworth filter\cite{butterworth} was chosen for filtering, as it exhibits a much smoother decrease in filter magnitude with increasing frequency, as opposed to higher-order filters. A lower order filter (first order in this case) would result in a gradual attenuation of noisy components in a signal, which will increase the allowable margin of error in an instance where the chosen cutoff frequency is non-optimal.\par

To determine the optimal cutoff frequency, an optimizer was developed, where its objective function was expressed by,

\begin{eqnarray}
\label{eq:objectivefn}
    max _\omega J(\omega) =  a \cdot J_{R}(\omega) + b \cdot J_{PSD}(\omega)
\end{eqnarray}

where $J_{R}(\omega)$ is the normalized cross-correlation fitness function, $J_{PSD}(\omega)$ is the normalized Power Spectral Density (PSD) fitness function and $a$, $b$ are scaling constants. The choice of \eref{eq:objectivefn} as the objective function, was subject to the easy manipulation and tuneable nature of this function, by changing the parameters $a$ and $b$. \par

Ideally, in a region with a high testing frequency and consistent testing patterns (i.e., a much smoother epi-curve), the filtering should be minimal. This is because a relatively larger percentage of the useful information will be encapsulated in the original signal itself, in contrast to much noisier signals which correspond to \covid\ epi-curves in regions with sporadic and limited testing practices. In technical terms, smoother epi-curves will exhibit a high Signal-to-Noise ratio (SNR). This retainment of information is quantified using $J_{R}(\omega)$, which is denoted by,

\begin{eqnarray}
    J_{R}(\omega) =  E[X_{initial} \cdot X_{filtered}(\omega)]
\end{eqnarray}
where $E$ is the expectation operator, $X_{initial}$ and $X_{filtered}(\omega)$ correspond to the original unfiltered signal and the filtered signal at a cutoff frequency of $\omega$ respectively. Here, $X_{initial}$ is the term used to denote the entirety of the initial signal, expressed by,
\begin{eqnarray}
    X_{initial} =  X_{signal} + X_{noise}
\end{eqnarray}
where $X_{signal}$ is the 'true' nature of the epi-curve (which needs to be estimated) and $X_{noise}$ is the fluctuating components caused by measurement noise, a result of irregular testing patterns. \par

To demonstrate the need for a $J_{PSD}(\omega)$ in addition to $J_{R}(\omega)$ in \eref{eq:objectivefn}, let us first consider a case where $X_{noise}$ is purely white. That is, the testing patterns are completely random and show no deterministic nature at all. Here, optimal filtering can be achieved by completely removing white noise, as the cross-correlation component (retainment of information) between two signals do not change with the addition or removal of white noise, hence, $J_{R}(\omega)$ will be equal to $1$ and the initial and filtered signals will be perfectly correlated. However, as previously observed in \fref{fig:autocorr}, external factors do contribute to testing patters of a region, thereby making $X_{noise}$ deterministic to an extent. As this is the case, only removing the white noise components will not sufficiently remove $X_{noise}$; the main objective of the smoothing algorithm. It will result in the output signal being under-filtered, due to retainment of some parts of $X_{noise}$, which can be considered 'coloured' noise. This implies that achieving $J_{R}(\omega)=1$ does not result in optimal filtering most of the time, and in some instances, the above condition will not be satisfied unless the filtered signal directly overlaps the original signal (i.e., no filtering). Hence, the need arises for another term $J_{PSD}(\omega)$, that attempts to further attenuate coloured noisy components of the signal.\par 

\begin{figure}[!ht]
    \includegraphics[width=0.6\textwidth]{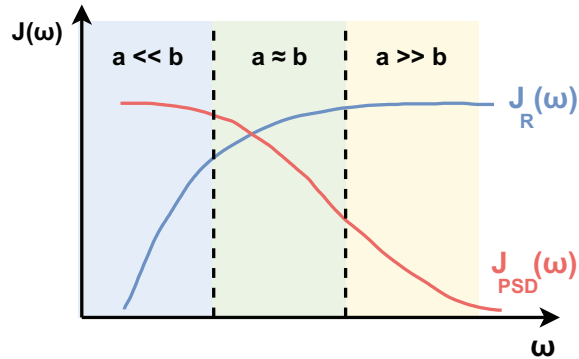}
    \caption{\bf Dynamics of $J(\omega)$ under multiple $a$ and $b$ settings}
    \label{fig:obj_extremes}
\end{figure}

$J_{PSD}(\omega)$ is computed by spectral analysis of the given noisy signal. The Power Spectral Density (PSD) of a signal is the measure of energy of that signal as a function of its frequency components. In other words, how much of the signal consists of high frequency components, and how much of it consists of low frequency components. High PSD values towards the latter end of the frequency spectrum suggests that the signal contains a large number of high-frequency components. The use of a low-pass filter results in an increased difference between the PSDs before and after filtering towards the higher frequency bands, which is used to compute the PSD fitness function given by,
\begin{eqnarray}
    J_{PSD}(\omega) =  \sum_{i=1}^{N}[g(i) \cdot (PSD_{initial}(i) - PSD_{filtered}(i,\omega))]
\end{eqnarray}
where $PSD_{initial}$ and $PSD_{filtered}$ are the N-point discretized PSDs of the original and filtered signals respectively, and $g$ is a monotonically increasing function. Due to its monotonically increasing nature, $g$ serves to provide more emphasis on the differences of PSDs towards the latter part of the frequency spectrum. Hence, $J_{PSD}(\omega)$ increases as the PSD difference in high-frequency bands increases. In this study, $g$ is a ramp function denoted by $g(i) = i$, for all $0\leq i\leq N$. The scaling constants, $a$ and $b$ used to tune $J(\omega)$, determine whether the overall objective function \eref{eq:objectivefn} relies more on noise reduction or information preservation. An illustration of its dynamics under multiple $a$ and $b$ settings is shown in \fref{fig:obj_extremes}.\par

As observed in the figure, when $a$ is significantly larger than $b$ (yellow/far right section),  $J_{R}(\omega)$ takes higher prominence in \eref{eq:objectivefn}, hence the final cutoff frequency will be higher, thereby focusing more on information preservation. This will possibly retain some $X_{noise}$ components in the filtered signal, resulting in under-filtering. In contrast, when $a<<b$ (blue/far left section), the objective function will be tuned such that a large chunk of high frequency components is removed by setting a very low cutoff frequency, thereby exposing the risk of over-filtering, due to removal of important signal information from $X_{signal}$ in addition to $X_{noise}$. Therefore, to eliminate possible effects of under-filtering and over-filtering the signal, $a$ and $b$ must be pre-set such that it falls in the green/middle section denoted by \fref{fig:obj_extremes}. It was found out through trial and error, that an $a/b$ ratio between $1.00$ and $1.50$ was suitable for this study. This ratio will essentially serve as a constraint in the optimization process, where the actual optimization will be carried out in \eref{eq:objectivefn}, within the bounds of $a$ and $b$. It should be noted that a poor choice of these parameters will result in all epi-curves either being over-filtered or under-filtered, regardless of the optimization procedure.  \par

\subsubsection*{Validation using \covid\ alert level analysis}

An analysis of alert levels computed using smoothed and non-smoothed data was carried out to validate the proposed smoothing technique. As previously mentioned in relation to testing regions, computation of alert levels in a number of districts in Sri Lanka was carried, out using two methods as described in the Introduction section which are based on \cite{alert_USA}. These alert level systems were redefined as the Low-Inertial alert level and the High-Inertial alert level. This is because one method's response to fluctuating data is imminent, and the other hardly responds to fluctuating data. The specific use of districts in Sri Lanka was due to their high volatility, as smoother regions in general are less affected by smoothing, they emulated the natural high-inertial tendency. More details regarding this will be noted in the later Results section. \par

According to \cite{alert_USA}, three alert level systems are introduced for disease situation, health care system and disease control respectively. The number of daily cases per one million population (i.e., daily incidence) was defined as the "disease situation". alert level thresholds were assigned as levels 1, 2, 3 and 4 if the daily case incidence was lower than 10, between 10 and 19, between 20 and 40, above 40 respectively. The conditions met to increase or decrease a level were set such that the nature of the alert level shows high inertia in one case, and low inertia in the other. To obtain high inertia, a level was increased if an upper threshold was met for 7 consecutive days and the alert level was decreased if a lower threshold was met for 14 consecutive days. In contrast, the low inertial alert level was obtained by changing the alert level instantaneously based on the daily case incidence. To ensure consistency in analysis, threshold values of both the low and high inertial alert levels were kept equal.  \par

Unlike forecasting, the 'goodness' of the behaviour of an alert level cannot be directly computed. Therefore, a qualitative analysis was carried out comparing how well the behaviour of an alert level relates to a real-world situation. It is widely known that alert levels in general, exhibit a high inertial nature. That is, once it changes from one state to another, it tends to stay in that state for at least a number of days prior to moving up or down a level. Therefore for the low inertial case, the effect of smoothing was quantified by the reduction in the number of 'spikes' in time, where a spike is defined as the instance where a level would constantly change within a span of three consecutive days.\par

However, taking into account the high inertial alert level, if the number of cases per day were to keep fluctuating, levels would not change due to the number of daily \covid\ cases would not consistently remain above or below a certain threshold for a consecutive number of days. Hence, the effect of smoothing on the high-inertial alert level was evaluated by observing the frequency of changes in the alert level over time, as the lack of fluctuations would result in a more realistic representation of alert level behaviour. Essentially, the evaluation of the two aforementioned alert levels is carried out by analysing how each alert level tends to display traits contrary to its main attribute. In other words, how high inertial characteristics are shown in the low inertial alert level and vice-versa. \par 

In addition to validation using \covid\ alert levels, the proposed smoothing algorithm was compared with N-day averaging, the most commonly used de-noiser for \covid\ related time-series models. \par

\subsection*{\covid\ forecasting model}

\subsubsection*{Data pre-processing and train-test splits}

\begin{figure}[!ht]
     \includegraphics[width=0.6\linewidth]{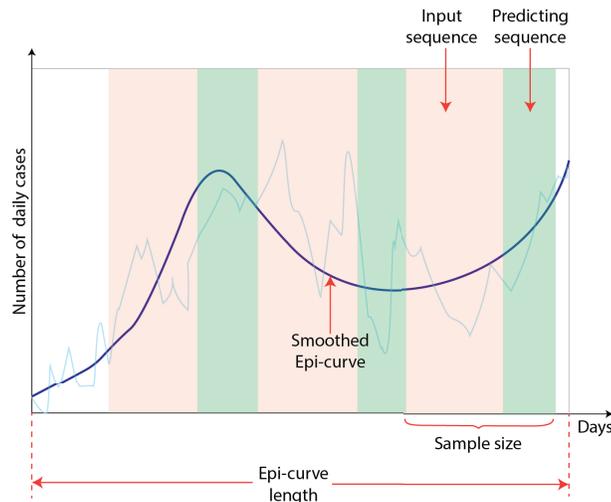}
    \caption{\bf A sample extracted from an epi-curve for the training dataset.}
    \label{fig:sample_from_epi-curve}
\end{figure}

To obtain optimal performance from predictive models, data was pre-processed through normalisation \cite{normalisation}. As some highly populated regions displayed a large number of cases and vice-versa, each region was 0-1 normalized, hence allowing the predictive model to train only based on the shape and trends of each epi-curve. These normalized data were then smoothed as will be discussed further under the Proposed optimal smoothing technique. Training samples were obtained by extracting several samples from each normalized and smoothed epi-curve. Each sample was then split into two parts, the input sequence of daily new cases and the sequence of daily new cases that needs to be predicted using the input sequence, as illustrated in \fref{fig:sample_from_epi-curve}.

As observed in \fref{fig:sample_from_epi-curve}, the size of a sample is the sum of the number of prediction input and output days.  Hence, to ensure no overlap and mutual exclusivity between samples, the maximum number of samples that can be extracted from an epi-curve is the time period of an existing epi-curve divided by the sample size. Therefore, the total number of training samples from all datasets will be this maximum number for each epi-curve, multiplied by the total number of epi-curves per region, times the number of training regions. All training samples were extracted from epi-curve data before 1st of March 2021, and the trained models were tested using epi-curve data after 1st of March 2021.\par

\subsubsection*{Neural Network architecture, Adaptive loss function and optimizer}

A long-short-term memory network (LSTM) \cite{LSTM}, which is widely used for time-series predictions was designed as the proposed NN model. A main feature of LSTM is that it mainly relies on selective storage of previous time-series data, thus relieving memory constraints. The trends visually observed in previous epidemiological data and the ability to selectively store information led to LSTMs being the choice of network for prediction, over more primitive algorithms such as Dense Neural networks. In addition, the abundant use of LSTMs in \covid\ forecasting literature led to the clear and obvious choice of LSTMs as the baseline forecasting model.\par

\begin{figure}[!ht]
     \includegraphics[width=0.6\linewidth]{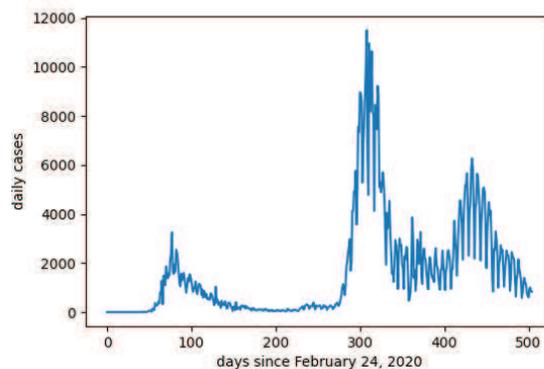}
    \caption{\bf Daily new \covid\ cases in Lombardy, Italy}
    Note that almost half the values in this epi-curve are close to zero.
    \label{fig:data_distribution}
\end{figure}

Epi-curves that belong to case studies that contain longer periods of data, display a multi-modal nature. Hence, they contain a large number of zeroes. That is, except during the periods of \covid\ waves, daily case incidences hardly exist. This larger distribution of zeros and values close to zero produces epi-curves that belong to regions such as Lombardy in Italy as shown in \fref{fig:data_distribution}, that contribute to an unbalanced dataset, resulting in an inefficient training process for forecasting models. This is due to the predictive model seeing a large number of zeroes whilst training, thereby assuming a high probability of zeros in its prediction. \par

To tackle this condition, two common methods exist; under-sampling and adaptive loss functions \cite{cost_vs_undersampling}. Under-sampling extracts several samples from the original dataset to create a new dataset, randomly or by a pre-determined sampling function. However, under-sampling reduces the number of training samples, and is inefficient in the case of \covid, where the size of training data is limited. On the other hand, adaptive loss functions are used in the model training process. They consider the position of each training sample in the whole distribution and amplify/attenuate the loss depending on the rarity/abundance of that sample. A density-based adaptive loss function was proposed, derived from the standard loss metric; Mean Squared Error (MSE), and this function is expressed by,

\begin{eqnarray}
    \mathcal{L} = 1/n \sum_{i\in\mathcal{B}} {\sum_{t}\frac{\left(y_{true}(t)-y_{pred}(t)\right)^2}{10\log^2 f(i(t))}}
    \label{eq:dba_loss_f}
\end{eqnarray}

where $\mathcal{B}$ is the set of samples in a batch, $n$ is the batch size, and $t$ is a particular day from the predicted days. The function $f$ returns the new daily cases $i(t)$ count in the dataset. In simple terms, \eref{eq:dba_loss_f}  implies that lesser occurring samples would display an amplified error metric $\mathcal{L}$ compared to the actual training error, and vise-versa.\par

Among NN optimizers such as Stochastic Gradient Descent (SGD), RMSprop, and Adam optimizers, Adam optimizer was chosen due to its consistent convergence towards an optimum solution \cite{adam}. \par

\subsubsection*{Evaluation methods}

Several methods were implemented to evaluate the proposed forecasting algorithm, both qualitatively and quantitatively. First, visual observation was carried out for the predicted vs observed samples to analyze the effect of smoothing and under-sampling. Then, the overall prediction accuracy was calculated and analyzed with and without the smoothing and under-sampling techniques. As listed in \tref{tab:methods_of_evaluation}, between smoothed/raw training data and types of loss functions, four distinct models exist. The accuracies of these four models were initially calculated for each model to forecast daily cases 10 days into the future, given 50-day previous data. These accuracies were then analyzed to choose the best combination of smoothing techniques and loss functions, and the corresponding model was chosen as the optimal model. To quantify all prediction errors, the Mean Absolute Error (MAE) metric was used. \par

\begin{table}[!ht]
\centering
\caption{\bf Methods used for evaluation}
    \begin{tabular}{|l+l|l|l|}
    \hline
    \bf Method & \bf  Model name      & \bf Training Data     & \bf Loss function          \\ \thickhline
    $A$      & LSTM-R-N   & Raw               & Standard MSE                    \\ \hline
    $B$      & LSTM-R-L   & Raw               & Proposed function   \\ \hline
    $C$      & LSTM-F-N   & Smoothed          & Standard MSE                    \\ \hline
    $D$      & LSTM-F-L   & Smoothed          & Proposed function   \\ \hline
    \end{tabular}

\label{tab:methods_of_evaluation}
\end{table}

All code was written in Python 3.8. Tensorflow 2.4 was used as the primary Machine Learning tool, in addition to the conventional data science libraries in Python 3.8.\par

\section*{Results and Discussion}
\subsection*{Optimized smoothing technique}

\begin{figure}[!ht]
    \includegraphics[width=\textwidth]{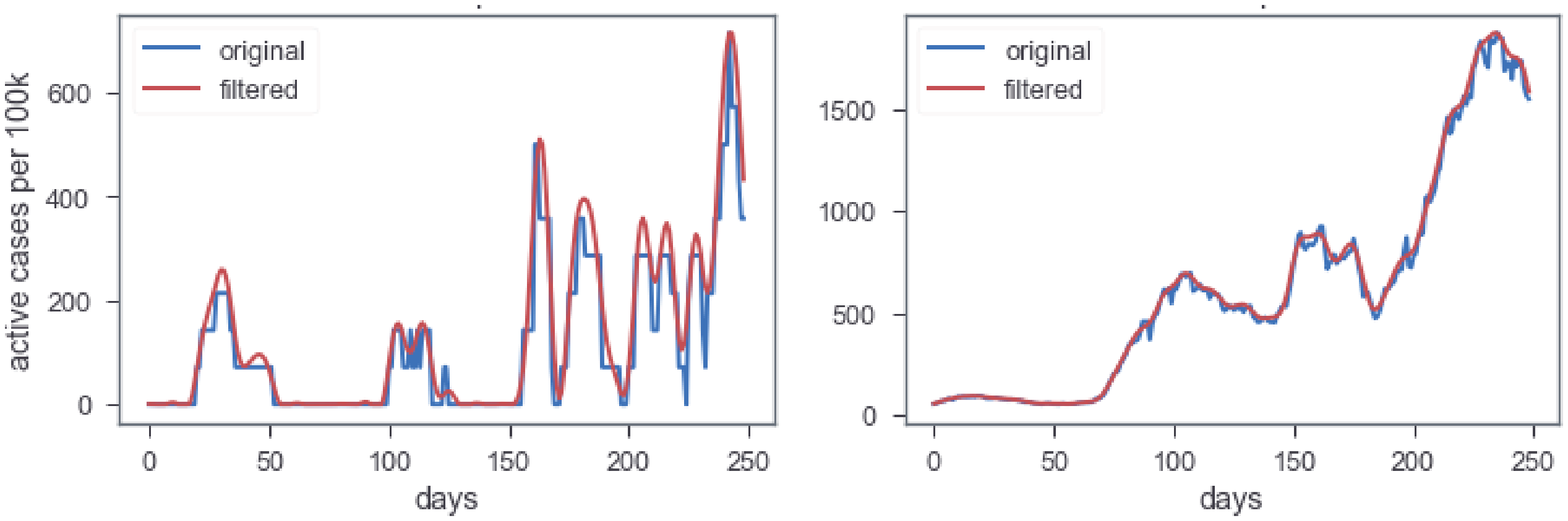}
    \caption{\bf Filtered signal with high cutoff frequency}
    (a)Cottle County. (b)Lubbock County.
    \label{fig:underfiltering}
\end{figure}
\begin{figure}[!ht]
    \includegraphics[width=\textwidth]{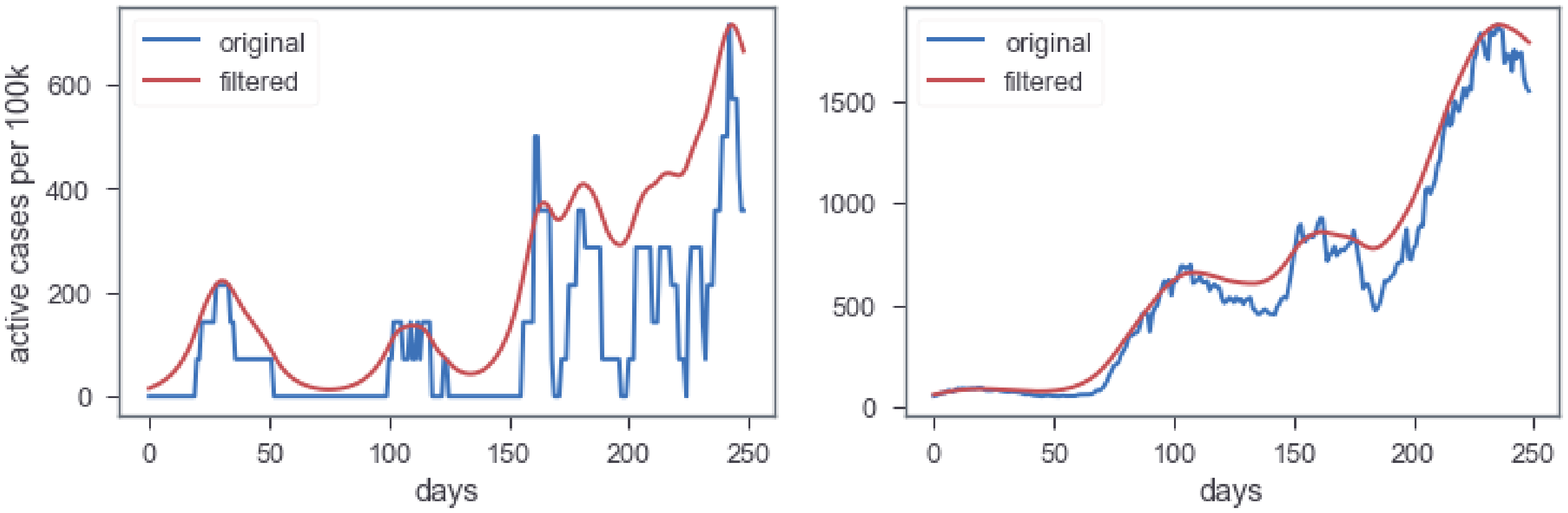}
    \caption{\bf Filtered signal with low cutoff frequency}
    (a)Cottle County. (b)Lubbock County.
    \label{fig:overfiltering}
\end{figure}

To reinforce the necessity of an optimized smoothed algorithm, epi-curves of each county were smoothed using LPFs with two distinct high and low cutoff frequencies. Taking into account the two counties of Texas mentioned in the epi-curve analysis, the smoothed signals using these high and low cutoff frequencies of these counties; Cottle and Lubbock county, are shown in \fref{fig:underfiltering} and \fref{fig:overfiltering} respectively.\par

As observed in \fref{fig:underfiltering}, although the signal for Lubbock county that is smoothed using a high-cutoff LPF is ideal, Cottle county displays a noisy curve, similar to the original signal. This is a result of under-filtering, where the high-frequency components still exist in the smoothed signal.\par
In contrast, excessively lowering the cutoff frequency to account for noisy signals results in over-filtering, where a loss of information is observed from less noisy initial signals such as Lubbock county, as shown in \fref{fig:overfiltering}.\par

\begin{figure}[!ht]
     \includegraphics[width=\textwidth]{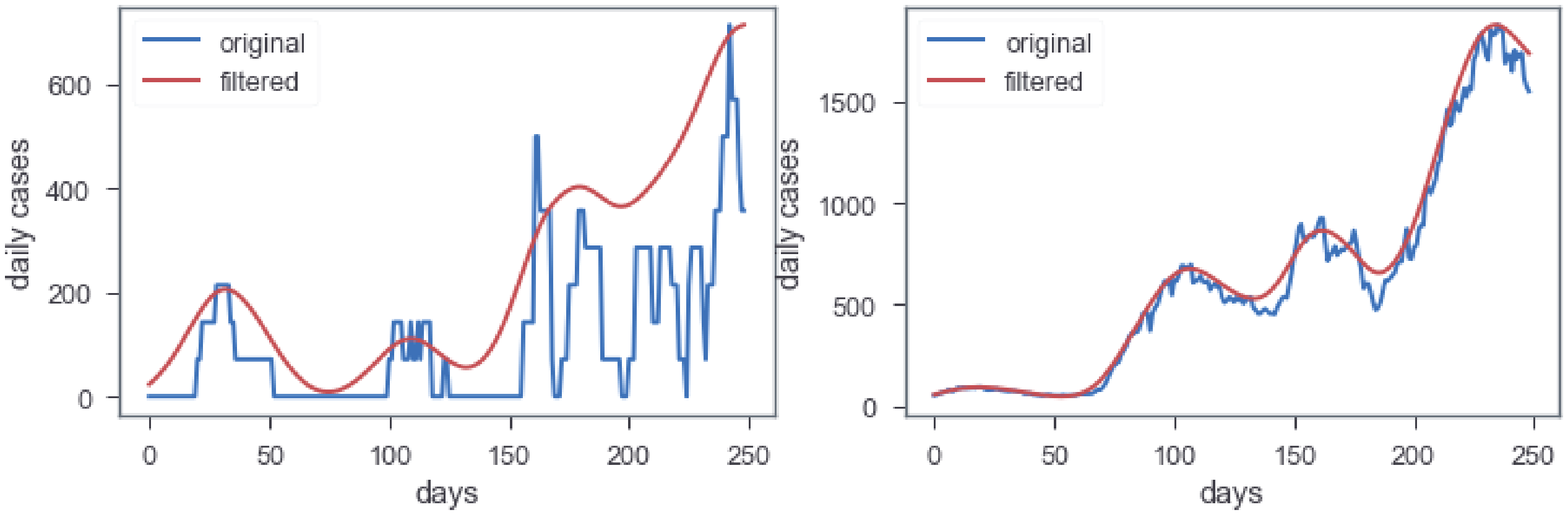}
    \caption{\bf Filtered signal with optimized cutoff frequencies}
    (a)Cottle County. (b)Lubbock County.
    \label{fig:optimal}
\end{figure}

The proposed optimized filter addresses the under-filtering and over-filtering problem, as illustrated in \fref{fig:optimal}.\par

From \fref{fig:optimal}, it is observed that the noisy nature of data is removed in Cottle county, whilst preserving nature of the curve in Lubbock county.\par

\begin{figure}[!ht]
    \centering
     \includegraphics[width=\textwidth]{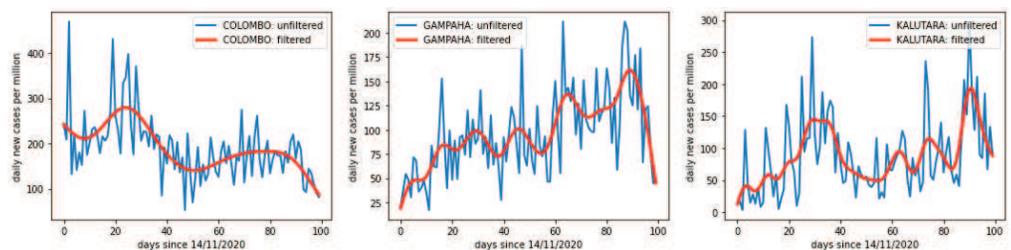}
    \caption{\bf Original vs filtered daily new cases for top 3 \covid\ affected districts in Sri Lanka}
    \label{fig:filterSL}
\end{figure}

In addition to counties of Texas, the daily new case data for districts in Sri Lanka was smoothed using the proposed optimized LPF. The original vs smoothed daily case data for the 3 most affected districts in Sri Lanka is shown in \fref{fig:filterSL}.\par

\begin{figure}[!ht]
     \includegraphics[width=0.6\textwidth]{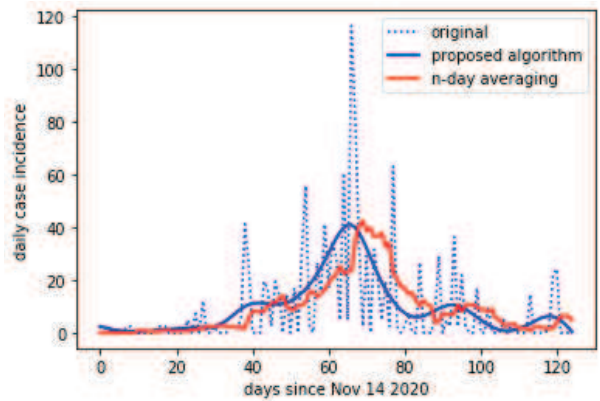}
    \caption{\bf Comparison of N-day averaging and proposed smoothing algorithm}
    \label{fig:smooth_nday}
\end{figure}

Another important feature of the proposed smoothing algorithm is the removal of time delay that is otherwise caused by existing smoothing methods for \covid\ such as n-day averaging. Consider the epi-curve of a district in Sri Lanka smoothed using the proposed algorithm and n-day averaging. As observed in \fref{fig:smooth_nday}, there exists a time delay between the original and smoothed signal, when smoothed using n-day averaging. This phenomenon is common for all smoothing algorithms evaluated in the time domain. In contrast, this delay does not affect smoothed signals evaluated using frequency domain analysis, as they are time invariant. It can also be observed that although n-day averaging reduces noise to an extent, there always exists a non-determinant nature, as opposed to the proposed smoothing algorithm.\par

\subsubsection*{Validation using \covid\ alert level analysis}

\begin{figure}[!ht]
    \includegraphics[width=\textwidth]{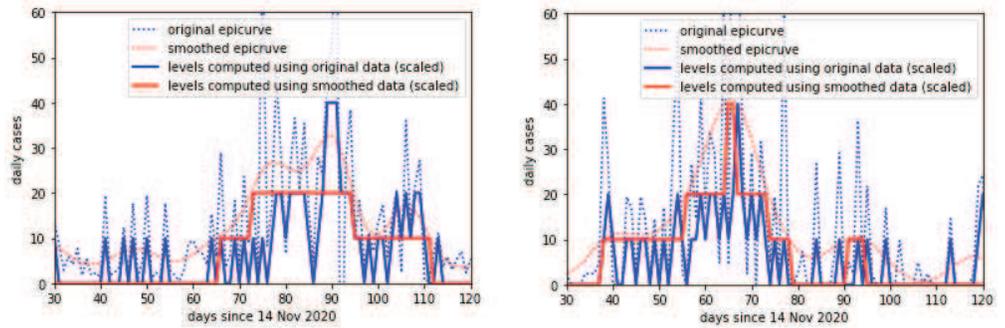}
    \caption{\bf The behaviour of low-inertial alert levels in districts of Sri Lanka.} (left) Kurunegala. (right) Trincomalee.
    \label{fig:alert_low_kur_tri}
\end{figure}

The optimal smoothing technique resulted in a significant reduction of 'spikes' in the low-inertial alert level, as shown in \fref{fig:alert_low_kur_tri}, where the low-inertial alert level for the districts of Kurunegala and Trincomalee in Sri Lanka was computed using original and smoothed data. The total number of spikes for all districts decreased from 676 to 4 over the 120-day period considered for all districts upon smoothing, which resulted in a much more realistic representation of \covid\ alert levels in time.\par

\begin{figure}[!ht]
     \includegraphics[width=\textwidth]{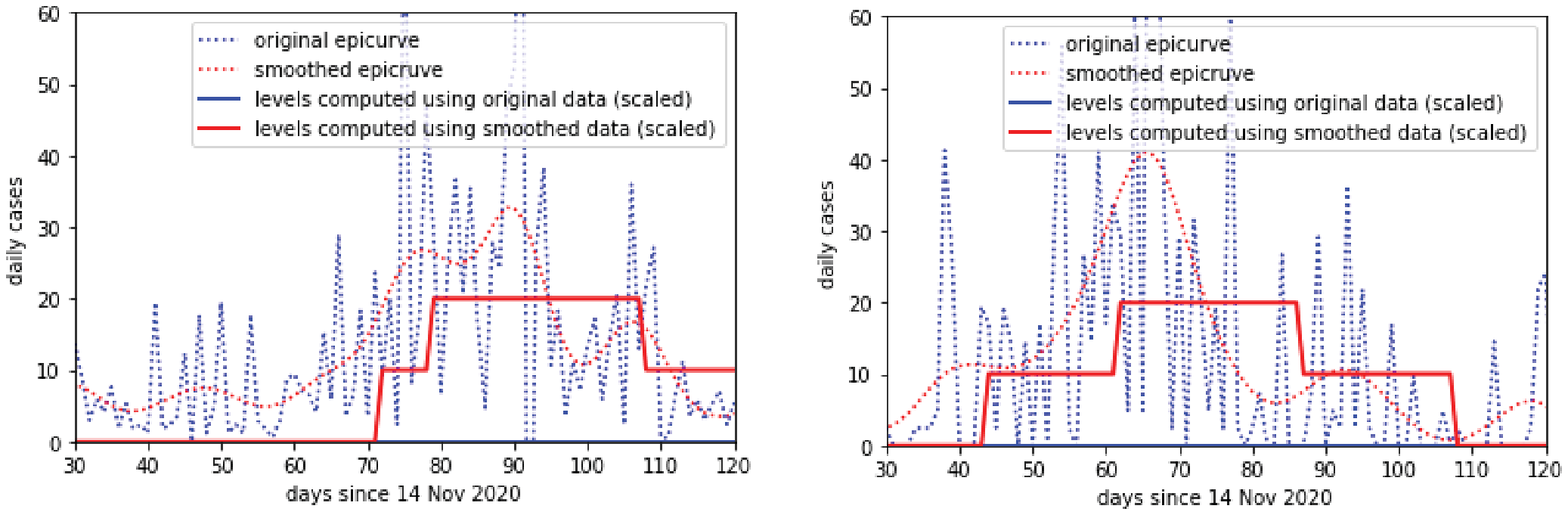}
    \caption{\bf Behaviour of high-inertial alert levels in disricts of Sri Lanka.}
    (left) Kurunegala. (right) Trincomalee.
    \label{fig:alert_high_kur_tri}
\end{figure}

Considering the high-inertial alert level, in contrast to districts with less fluctuating epi-curves which indicated little change in behaviour due to smoothing, districts corresponding to higher fluctuating epi-curves such as Kurunegala and Trincomalee showed significant changes in behaviour.  This is observed in figure \fref{fig:alert_high_kur_tri}, where these districts which were previously at a constant alert level throughout the 120-day period, now display trends that align more towards the real disease situation (the reported number of cases). \par

To further affirm the advantage of the reduction of 'spikes', consider the alert levels of Kurenegala in \fref{fig:alert_low_kur_tri}. When comparing the smoothed and original low-inertial alert levels during the 40-60 day period since 14 November 2020, it is observable that there exists a fair number of cases during that period. This corresponds to an increase in the alert level computed from original data. However, the smoothed alert level does not change from zero as the average value is less than the threshold at which the level would increase. This explains that the alert level computed using original data is due to the backlog of observed cases reported on each day, where cases are reported every 2-3 days. Hence, policy makers would not need to react hastily due to the reported number of cases.\par

From \fref{fig:alert_low_kur_tri} and \fref{fig:alert_high_kur_tri} it is observable that the similarity between the low-inertial and high-inertial alert levels are much higher when computed using smoothed epi-curves, as opposed to when computed using original epi-curves. This allows the low-inertial alert level, which operates on real-time data, to be a useful indicator of the current disease situation. In other words, the smooth nature of the high-inertial alert level is encapsulated in the low-inertial alert level, when computed using smoothed epi-curves. This is advantageous for policy makers as they do not have to resort to analysing the high inertial alert level which carries an inherent lag in time. Also, issues due to backlogs which can naturally occur in times of pandemics with the associated heavy caseloads are auto-corrected by the proposed mechanic preventing hasty conclusions to noisy data.\par

\subsection*{Forecasting daily new \covid\ cases}

\begin{figure}[htb]
    \includegraphics[width=\linewidth]{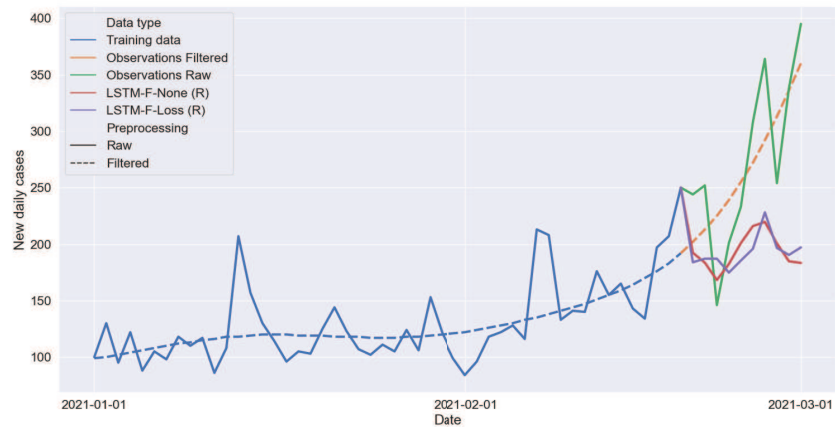}
    \caption{\bf An example prediction for the Hokkaido region in Japan from the model trained using raw data.} Models that were trained using the generalized training strategy are in red and models trained using the proposed loss function are shown in purple.
\label{fig:pred_raw}
\end{figure}
\begin{figure}[htb]
    \includegraphics[width=\linewidth]{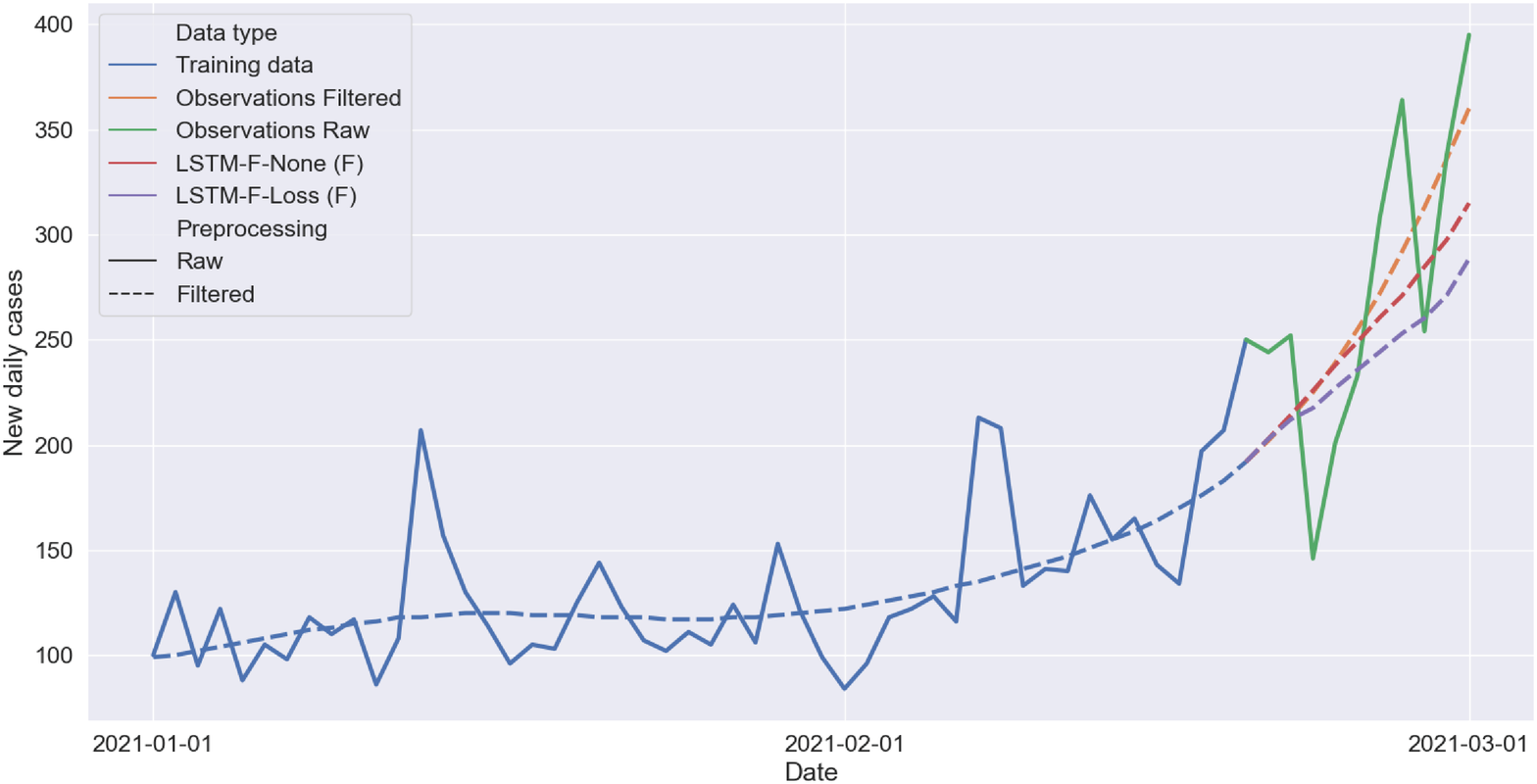}
    \caption{\bf An example prediction for the Hokkaido region in Japan from the model trained using smoothed data.}Models that were trained using the generalized training strategy are in red and models trained using the proposed loss function are shown in purple.
\label{fig:pred_fil}
\end{figure}

Daily new \covid\ cases in testing regions were predicted 10 days ahead, given 50-day previous data using the LSTM-based NN model, and was evaluated using Methods A, B, C and D as mentioned in \tref{tab:methods_of_evaluation}. The prediction of daily cases using these methods for Hokkaido (a prefecture of Japan) is visually illustrated in \fref{fig:pred_raw} for the model trained with raw data, and \fref{fig:pred_fil} for the model trained using smoothed data. \par

It is observable that the predicted data points of the model trained using optimally smoothed data (LSTM-F) (\fref{fig:pred_fil}) show fewer fluctuations than the model trained using raw data (LSTM-R) (\fref{fig:pred_raw}). In addition, the output from LSTM-F when compared with raw (un-smoothed) data also shows an increase in accuracy, in comparison to the errors obtained by outputs of LSTM-R compared with raw data. Although this increase cannot be clearly observed from the figure, \tref{tab:pred_results_all} shows an average decrease in mean error by 47\% when the model is trained using the adaptive loss function (Method D vs B) for raw to smoothed data inputs and 70\% when trained using the standard MSE loss function (Method C vs A) for raw to smoothed data inputs. The removal of noisy components (by smoothing) whilst training provides better prediction accuracies even when the smoothed output is compared with raw, noisy ground truth. This implies a robust training process, caused by the lack of meaningless data (noise) seen by the model during training, thus minimizing the possible instances of over-fitting.\par

\begin{table}[!ht]
\begin{adjustwidth}{-2.25in}{0in} 

\caption{\bf Prediction errors for 50-day input 10-day output forecasts}
\centering
\resizebox{\linewidth}{!}{%
\begin{tabular}{lcc|c|c|c|c|c|c|c|c|}
\cline{4-11}
 &
  \multicolumn{2}{l}{} &
  \multicolumn{8}{|c|}{\textbf{Ground truth dataset and error (MAE/cases)}} \\ \hline
\multicolumn{3}{|l|}{\textbf{Training data and method}} &
  \multicolumn{2}{c|}{\textbf{JPN}} &
  \multicolumn{2}{c|}{\textbf{LKA}} &
  \multicolumn{2}{c|}{\textbf{RUS}} &
  \multicolumn{2}{c|}{\textbf{NW}} \\ \hline
\multicolumn{1}{|l|}{\textbf{Data type}} &
  \multicolumn{1}{l|}{\textbf{Country}} &
  \multicolumn{1}{l|}{\textbf{Method}} &
  \textbf{Raw} &
  \textbf{Smooth} &
  \textbf{Raw} &
  \textbf{Smooth} &
  \textbf{Raw} &
  \textbf{Smooth} &
  \textbf{Raw} &
  \textbf{Smooth} \\ \hline
\multicolumn{1}{|l|}{} &
  \multicolumn{1}{c|}{} &
  A &
  \cellcolor[HTML]{FFCCC9}60.92 &
  \cellcolor[HTML]{FFCCC9}49.21 &
  \cellcolor[HTML]{FFCCC9}52.26 &
  \cellcolor[HTML]{FFCCC9}9.3 &
  \cellcolor[HTML]{FFCCC9}49.21 &
  \cellcolor[HTML]{FFCCC9}49.92 &
  \cellcolor[HTML]{FFCCC9}49.92 &
  \cellcolor[HTML]{FFCCC9}28.67 \\ \cline{3-11} 
\multicolumn{1}{|l|}{} &
  \multicolumn{1}{c|}{} &
  B &
  \cellcolor[HTML]{FFCCC9}45.84 &
  \cellcolor[HTML]{FFCCC9}34.01 &
  \cellcolor[HTML]{FFCCC9}23.96 &
  \cellcolor[HTML]{FFCCC9}7.55 &
  \cellcolor[HTML]{FFCCC9}34.01 &
  \cellcolor[HTML]{FFCCC9}20.26 &
  \cellcolor[HTML]{FFCCC9}20.26 &
  \cellcolor[HTML]{FFCCC9}23.68 \\ \cline{3-11} 
\multicolumn{1}{|l|}{} &
  \multicolumn{1}{c|}{} &
  C &
  \cellcolor[HTML]{9AFF99}37.02 &
  \cellcolor[HTML]{9AFF99}12.82 &
  \cellcolor[HTML]{9AFF99}10.37 &
  1.06 &
  \cellcolor[HTML]{9AFF99}12.82 &
  \cellcolor[HTML]{9AFF99}3.32 &
  3.32 &
  1.81 \\ \cline{3-11} 
\multicolumn{1}{|l|}{\multirow{-4}{*}{\begin{tabular}[c]{@{}l@{}}Selected \\ regions\\ for training\end{tabular}}} &
  \multicolumn{1}{c|}{\multirow{-4}{*}{\begin{tabular}[c]{@{}c@{}}US-TX-counties, NGA-states, \\ ITA-provinces, BGD-districts, \\ KAZ-provinces, KOR-cities, \\ DEU-states\end{tabular}}} &
  D &
  38.43 &
  14.24 &
  \cellcolor[HTML]{9AFF99}9.99 &
  \cellcolor[HTML]{9AFF99}0.94 &
  14.24 &
  \cellcolor[HTML]{9AFF99}2.74 &
  \cellcolor[HTML]{9AFF99}2.74 &
  \cellcolor[HTML]{9AFF99}1.46 \\ \hline
\multicolumn{1}{|l|}{} &
  \multicolumn{1}{c|}{} &
  C &
  \cellcolor[HTML]{96FFFB}46.02 &
  \cellcolor[HTML]{96FFFB}25.16 &
  \multicolumn{2}{c|}{} &
  \multicolumn{2}{c|}{} &
  \multicolumn{2}{c|}{} \\ \cline{3-5}
\multicolumn{1}{|l|}{} &
  \multicolumn{1}{c|}{\multirow{-2}{*}{JPN-prefectures}} &
  D &
  \cellcolor[HTML]{96FFFB}36.53 &
  \cellcolor[HTML]{96FFFB}13.55 &
  \multicolumn{2}{c|}{\multirow{-2}{*}{n/a}} &
  \multicolumn{2}{c|}{\multirow{-2}{*}{n/a}} &
  \multicolumn{2}{c|}{\multirow{-2}{*}{n/a}} \\ \cline{2-11} 
\multicolumn{1}{|l|}{} &
  \multicolumn{1}{c|}{} &
  C &
  \multicolumn{2}{c|}{} &
  \cellcolor[HTML]{96FFFB}14.68 &
  \cellcolor[HTML]{96FFFB}7.77 &
  \multicolumn{2}{c|}{} &
  \multicolumn{2}{c|}{} \\ \cline{3-3} \cline{6-7}
\multicolumn{1}{|l|}{} &
  \multicolumn{1}{c|}{\multirow{-2}{*}{LKA-districts}} &
  D &
  \multicolumn{2}{c|}{\multirow{-2}{*}{n/a}} &
  \cellcolor[HTML]{96FFFB}14.81 &
  \cellcolor[HTML]{96FFFB}8.38 &
  \multicolumn{2}{c|}{\multirow{-2}{*}{n/a}} &
  \multicolumn{2}{c|}{\multirow{-2}{*}{n/a}} \\ \cline{2-11} 
\multicolumn{1}{|l|}{} &
  \multicolumn{1}{c|}{} &
  C &
  \multicolumn{2}{c|}{} &
  \multicolumn{2}{c|}{} &
  \cellcolor[HTML]{96FFFB}14.45 &
  \cellcolor[HTML]{96FFFB}8.88 &
  \multicolumn{2}{c|}{} \\ \cline{3-3} \cline{8-9}
\multicolumn{1}{|l|}{} &
  \multicolumn{1}{c|}{\multirow{-2}{*}{RUS-cities}} &
  D &
  \multicolumn{2}{c|}{\multirow{-2}{*}{n/a}} &
  \multicolumn{2}{c|}{\multirow{-2}{*}{n/a}} &
  \cellcolor[HTML]{96FFFB}25.12 &
  \cellcolor[HTML]{96FFFB}19.72 &
  \multicolumn{2}{c|}{\multirow{-2}{*}{n/a}} \\ \cline{2-11} 
\multicolumn{1}{|l|}{} &
  \multicolumn{1}{c|}{} &
  C &
  \multicolumn{2}{c|}{} &
  \multicolumn{2}{c|}{} &
  \multicolumn{2}{c|}{} &
  \cellcolor[HTML]{96FFFB}39.85 &
  \cellcolor[HTML]{96FFFB}28.5 \\ \cline{3-3} \cline{10-11} 
\multicolumn{1}{|l|}{\multirow{-8}{*}{\begin{tabular}[c]{@{}l@{}}From test \\ regions\end{tabular}}} &
  \multicolumn{1}{c|}{\multirow{-2}{*}{NOR-states}} &
  D &
  \multicolumn{2}{c|}{\multirow{-2}{*}{n/a}} &
  \multicolumn{2}{c|}{\multirow{-2}{*}{n/a}} &
  \multicolumn{2}{c|}{\multirow{-2}{*}{n/a}} &
  \cellcolor[HTML]{96FFFB}45.88 &
  \cellcolor[HTML]{96FFFB}45.92 \\ \hline

\label{tab:pred_results_all}
\end{tabular}%
}
\end{adjustwidth}
\end{table}

On the other hand, comparing LSTM-F and LSTM-R outputs with pre-smoothed data as the ground truth shows a significant decrease in mean error, by approximately 77\% when the model was trained using the adaptive loss function (Method D vs B: for raw vs smoothed data inputs) and 86\% when trained using the standard MSE loss function (Method C vs A: for raw vs smoothed data inputs). In summary, the LSTM-based NN model trained on smoothed data (Method C and D) outperforms the model trained on raw data (Method A and B), under all testing circumstances considered above. \par

From \tref{tab:pred_results_all}, considering the models trained using unsmoothed (raw) data from all training datasets (highlighted in red), the model trained using the proposed loss function (Method B) performs approximately 40\% better when compared to the model trained using normal loss (Method A). This observation is more significant when models were trained using original data (non-smoothed) rather than smoothed data (Method D vs C).Hence, the proposed adaptive loss function by itself contributes to a standalone performance increase when used independently to the smoothing technique. This might be useful for policy makers who decide to utilize raw data as opposed to smoothed data due to high confidence in historical data. This is mostly applicable to countries with high test rates.\par

\begin{figure}[!ht]
    \includegraphics[width=\textwidth]{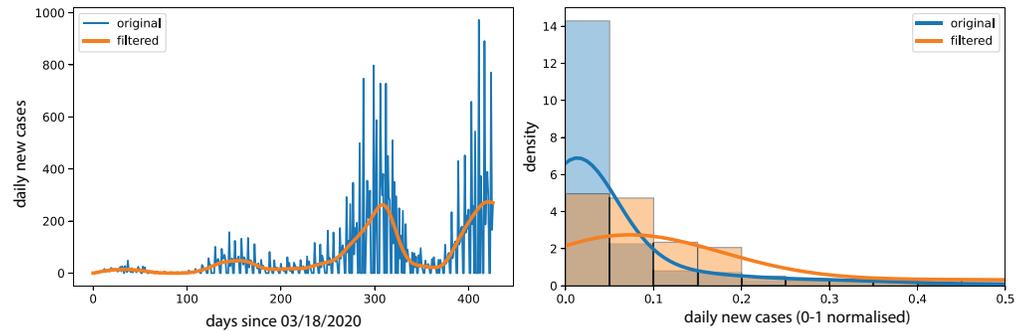}
    \caption{\bf Original and smoothed epi-curves and their density distributions in the region of Kyoto, Japan.}
    (left) Original and Smoothed epi-curves. (right) Density distributions.
    \label{fig:raw_fil_dist}
\end{figure}

The reason for the increase in accuracy of Method B compared with Method A is the highly unbalanced nature that exists in the original dataset due to the disproportionate number of recorded cases being zero or in its neighborhood, caused by unrecorded \covid\ cases in regions where testing is periodic and long periods between \covid\ waves in larger datasets. In the latter instance, the nature of smoothed epi-curves do not deviate from the original epi-curve. However, the number of unrecorded cases in-between days significantly reduces upon smoothing. A perfect example to explain this phenomenon would be the region of Kyoto in Japan, shown in \fref{fig:raw_fil_dist}, where its epi-curve contains a large number of unrecorded (zero) days prior to smoothing. During the training process, the proposed adaptive loss function based predictive model 'ignores' these zeros due to its large number of occurrences by attenuating the loss function, and provide more emphasis towards predicting the non-zero values in the epi-curve. \fref{fig:raw_fil_dist} also shows that the smoothed epi-curve discards these zero values, thereby altering the distribution of cases to be more balanced in comparison to the distribution of raw data. As this characteristic is consistent throughout most regions extracted from training datasets, there is no significant change in prediction accuracy of Method D compared to Method C, where both models corresponding to these methods were trained using smoothed data.\par

From \tref{tab:pred_results_all}, when the selected regions (in \tref{tab:case_study}) based on geographical location and epi-curve variation, were used for training (the proposed generalized training strategy), method D outperformed method C in 3 out of 4 test regions, except for Japan, where it showed a slightly lesser performance. However, it should be noted that both Method C and D when trained using all test regions produce almost similar results. Hence, the LSTM-based NN model corresponding to method D (LSTM-F-L) was chosen as the best model for prediction, based on the increase in performance shown by the adaptive loss function when trained on raw data. This will result in the predictive model being sustainable even in the instance of sub-optimal smoothing. Here, the adaptive loss function performs to a certain extent, a similar role performed by the smoothing algorithm. This is due to the effect of non-recorded cases between periodic testing days being nullified by the attenuation in the loss function and rectified by the smoothing algorithm. However, a drawback of the adaptive loss function is that it also attenuates its loss in the periods that exhibit actual zero \covid\ cases, especially between waves in longer epi-curves. These periods, as opposed to non-recorded cases between days, should be learnt by the LSTM model, as they correspond to the behaviour of the increase and decrease of \covid\ waves in time.\par

To highlight the importance of the proposed generalized training strategy that chooses a variety of regions to train the predictive model, daily new \covid\ cases in each test region (refer \tref{tab:case_study}) were predicted using models of the same predictive model architecture, but trained only using historical data from that region as is the current practice in most of the existing body of work. These prediction errors are tabulated in \tref{tab:pred_results_all} (in cyan), where \covid\ cases in Japan are predicted using historical \covid\ data from Japan, \covid\ cases in Russia predicted using previous Russian data, and so forth. For all test regions, the model trained using the selected generalized strategy (from \tref{tab:case_study}) yields higher accuracy than the models trained using their own regions, by an average of 30\%. The increase in accuracy is resulted by the addition of external datasets such as the selected regions, which helps encapsulate a larger number of epi-curve scenarios that could occur in a new region, that might have not already occurred in that region. \par

\section*{Conclusion}

In this paper, an end-to-end forecasting system towards localized predictions of daily new incidences of \covid\ is proposed. The forecasting system consists of an optimized smoothing algorithm to smoothen fluctuating \covid\ epi-curves that result from inconsistent testing strategies, along with an LSTM-based predictive model trained using a generalized training strategy, in addition to a density-based adaptive loss function whilst training to tackle unbalanced datasets that are the result of high zero values in their epi-curves. The optimized smoothing algorithm uses an objective function that maximizes the smoothness of the resulting epi-curve whilst minimizing its deviation from the original epi-curve. In other words, it attempts to strike an optimal balance between retention of the signal or information component of the epi-curve whilst removing uncorrelated as well as correlated noisy elements utilizing digital signal processing techniques. The density-based loss function is employed during the model training process, where it attenuates the training loss depending on the abundance of a training sample in the training dataset. This mitigates tendencies to predict zero or low values due to their natural high presence in practical epi-curve data. The LSTM-based predictive model was trained using epi-curves from a wide range of regions, chosen such that the regions are diverse in both geographic location and nature of epi-curve. This generalized training strategy deviates from the norm in the existing state-of-the-art to utilize local region historical data for prediction purposes in COVID19 research thus far. Hence, it enables epidemiological dynamics due to different geographical, demographic, climatic and even special event-based to be learned by the model for prediction.\par

To demonstrate the effectiveness of the proposed optimized smoothing algorithm, an alert level analysis was carried out, where \covid\ alert levels were computed using original and smoothed data. It was observed that a more realistic alert level is obtained when computed using smoothed data, whilst preserving its real-time nature. It was also observed that the proposed smoothening technique that utilizes frequency-domain tools does not result in the time-lag prevalent in standard N-day moving average methods used in practice. The smoothing algorithm was also validated using the LSTM-based predictive model, where smoothing resulted in a 60\% increase in prediction accuracy. The density-based loss function performed best when models were trained using raw datasets, exhibiting a performance increase of 40\% in contrast to models trained using the standard MSE loss functions. This is owing to the ability of the loss function to correctly identify occurrences where unrecorded \covid\ cases exist, in regions where periodic testing is prevalent. Although the adaptive loss function doesn't show a significant increase in performance when trained using smoothed data, it serves as a buffer in instances where the smoothing algorithm fails or performs sub-optimally. This increases the robustness of the predictive model. For example, when in cases where policy makers prefer raw data due to high confidence in them in situations where high and consistent levels of testing take place. Essentially, smoothed data is a requirement for regions/countries with limited or inconsistent testing and for more affluent countries which performs higher rates of testing the adaptive loss mechanic offers more.  \par

The proposed generalized training strategy was validated, as models trained using these diverse representative datasets that formed a more wholistic basis for the training process, resulted in better accuracies when evaluated against models trained using historical data of individual test regions. The implementation possibilities of localized forecasting in regions where new \covid\ variants occur could be greatly increased by employing the proposed training strategy. For example, dynamics of the delta variant  that has already largely occured in some regions of the world can be learnt using this strategy, allowing accurate forecasts in regions where this variant has just arrived.  \par

\section*{Author Contribution}

UM, HW, RP, JH, SS, GJ, RG, VH, PE, JE, AR, and SD conceptualized 
the study and designed the methodology. RG, VH, PE, JE, AR, and SD supervised the project. UM, HW, RP, JH, SS, GJ carried out the literature 
search. UM, HW, RP carried out the methodology and wrote the draft. JH, SS, GJ, RG, VH, PE, JE, AR reviewed and edited the draft. UM, HW, RP, JH, SS, GJ, RG, VH, PE, JE, AR, and SD had access to the data, read the final manuscript and approved it.

\section*{Conflict of Interest}

The authors have no conflicts of interest to declare.




%
%
%


\begin{thebibliography}{10}




\bibitem{JHU_dashboard}
Dong E, Du H, Gardner L.
\newblock{An interactive web-based dashboard to track COVID-19 in realtime}.
\newblock The Lancet Infectious Diseases. 2020;20(5):533–-534.

\bibitem{travel_restrictions}
Chinazzi M, Davis JT, Ajelli M, Gioannini C, Litvinova M, Merler S, et al.
\newblock{The effect of travel restrictions on the spread of the 2019 novel coronavirus (2019-nCoV) outbreak}.
\newblock Science. 2020;368(April):395--400.

\bibitem{lockdown}
Aleta A, Mart´ın-Corral D, Pastore y Piontti A, Ajelli M, Litvinova M, Chinazzi M, et al.
\newblock{Modelling the impact of testing, contact tracing and household quarantine on second waves of COVID-19}.
\newblock Nature Human Behaviour. 2020;4(9):964--971.

\bibitem{pone_mental_lockdown}
Gao J, Zheng P, Jia Y, Chen H, Mao Y, Chen S, et al.
\newblock{Mental health problems and social media exposure during COVID-19 outbreak}.
\newblock PLoS ONE. 2020;15(4):1--10.

\bibitem{eci_non-evidence}
Ioannidis J.
\newblock{Coronavirus disease 2019: The harms of exaggerated information and non-evidence-based measures.}.
\newblock European Journal of Clinical Investigation. 2020;50(4):1--5.


\bibitem{pone_heartrate_lockdown}
Bourdillon N, Yazdani S, Schmitt L, Millet GP.
\newblock{Effects of COVID-19 lockdown on heart rate variability}.
\newblock PLoS ONE. 2020;15(11 November):1--10.


\bibitem{pone_mental_NJ}
Kecojevic A, Basch CH, Sullivan M, Davi NK.
\newblock{The impact of the COVID-19 epidemic on mental health of undergraduate students in New Jersey, cross-sectional study}.
\newblock PLoS ONE. 2020;15(9 September):1--16.


\bibitem{pone_psychological_US}
Browning MHEM, Larson LR, Sharaievska I, Rigolon A, McAnirlin O, Mullenbach L, et al.
\newblock{Psychological impacts from COVID-19 among university students: Risk factors across seven states in the United States}.
\newblock PloS one. 2021;16(1):e0245327. 


\bibitem{pone_depression_BD}
Akhtarul Islam M, Barna SD, Raihan H, Nafiul Alam Khan M, Tanvir Hossain M.
\newblock{Depression and anxiety among university students during the COVID-19 pandemic in Bangladesh: A web-based cross-sectional survey}.
\newblock PLoS ONE. 2020;15(8 August):1--12.


\bibitem{pone_edu_spain}
Gonzalez T, De la Rubia MA, Hincz KP, Comas-Lopez M, Subirats L, Fort S, et al.
\newblock{Influence of COVID-19 confinement on students’ performance in higher education}.
\newblock PLoS ONE. 2020;15(10 October):1--23.

\bibitem{small_businesses}
Bartik AW, Bertrand M, Cullen Z, Glaeser EL, Luca M, Stanton C.
\newblock{The impact of COVID-19
on small business outcomes and expectations}.
\newblock Proceedings of the National Academy of Sciences of the United States of America. 2020;117(30):17656--17666.

\bibitem{optimal_decisions}
Gombos K, Herczeg R, Eross B, Kov´acs SZ, Uzzoli A, Nagy T, et al.
\newblock{Translating Scientific Knowledge to Government Decision Makers Has Crucial Importance in the Management of the COVID-19 Pandemic}.
\newblock Population Health Management. 2021;24(1):35--45.

\bibitem{dzau2020strategy}
Dzau VJ, Balatbat C.
\newblock{Strategy, coordinated implementation, and sustainable financing
needed for COVID-19 innovations}.
\newblock The Lancet. 2020;396(10261):1469--1471.

\bibitem{alert_USA}
Alert-Level Systems for COVID-19; 2021.
\newblock{Available from:https://preventepidemics.org/covid19/resources/levels/\#designing-and-implementing-an-alert-level-system}.


\bibitem{alert_CDC}
How CDC Determines the Level for COVID-19 Travel Health Notices; 2021.
\newblock{Available from: https://www.cdc.gov/coronavirus/2019-ncov/travelers/how-level-is-determined.html}.


\bibitem{weiss2013sir}
weiss2013sir
\newblock{The SIR model and the foundations of public health}.
\newblock Materials matematics. 2013; p. 0001--17.

\bibitem{zeroual2020deep}
Zeroual A, Harrou F, Dairi A, Sun Y.
\newblock{Deep learning methods for forecasting COVID-19 time-Series data: A Comparative study}.
\newblock Chaos, Solitons \& Fractals. 2020;140:110121

\bibitem{AI_forecasting_review}
Shinde GR, Kalamkar AB, Mahalle PN, Dey N, Chaki J, Hassanien AE.
\newblock{Forecasting Models for Coronavirus Disease (COVID-19): A Survey of the State-of-the-Art}.
\newblock SN Computer Science. 2020;1(4):197.

\bibitem{jayatilaka2020use}
Jayatilaka GC, Hassan J, Marikkar U, Perera R, Sritharan S, Weligampola H, et al.
\newblock{Use of Artificial Intelligence on spatio-temporal data to generate insights during COVID-19 pandemic: A Review}
\newblock MedRxiv. 2020.


\bibitem{IEEE_TAI}
Islam M, Inan T, Rafi S, Akter S, Sarker IH, Islam A.
\newblock{A Systematic Review on the Use of AI and ML for Fighting the COVID-19 Pandemic}
\newblock IEEE Transactions on Artificial Intelligence. 2020;1(03):258--270.


\bibitem{pone_canada}
Chen LP, Zhang Q, Yi GY, He W.
\newblock{Model-based forecasting for Canadian COVID-19 data}
\newblock Plos One. 2021;16(1):e0244536.



\bibitem{pone_pakistan}
Chen LP, Zhang Q, Yi GY, He W, Ali M, Khan DM, et al. 
\newblock{Forecasting COVID-19 in Pakistan}
\newblock PLoS ONE. 2020;15(11 November):e0244536. 


\bibitem{forecast_brazil}
Ribeiro MHDM, da Silva RG, Mariani VC, Coelho LdS.
\newblock{Short-term forecasting COVID-19 cumulative confirmed cases: Perspectives for Brazil}
\newblock Chaos, Solitons and Fractals. 2020;135.


\bibitem{forecast_EU}
Kirbas, Sozen A, Tuncer D, Kazancıoglu Fikret S.
\newblock{Comparative analysis and
forecasting of COVID-19 cases in various European countries with ARIMA, NARNN and
LSTM approaches}
\newblock Chaos, Solitons and Fractals. 2020;138.


\bibitem{forecast_india}
Sarkar K, Khajanchi S, Nieto JJ.
\newblock{Modeling and forecasting the COVID-19 pandemic in India}
\newblock Chaos, Solitons and Fractals. 2020;139:110049.


\bibitem{pone_SIR_srilanka}
Ediriweera DS, de Silva NR, Malavige GN, de Silva HJ.
\newblock{An epidemiological model to aid decision-making for COVID-19 control in Sri Lanka}
\newblock PLoS ONE. 2020;15(8 August):1--10.


\bibitem{failed_forecasting}
Ioannidis JPA, Cripps S, Tanner MA.
\newblock{Forecasting for COVID-19 has failed}
\newblock International Journal of Forecasting. 2020.


\bibitem{MA_filters}

\newblock{Stock-market patterns and financial analysis:Methodologiva suggestions}.
\newblock The Journal of Finance. 1959;14(1):1--10.


\bibitem{LPF}
Kaiser JF, Reed WA.
\newblock{Data smoothing using low-pass digital filters}.
\newblock Review of Scientific Instruments. 1977;48(11):1447--1457.


\bibitem{cooper2020sir}
Cooper I, Mondal A, Antonopoulos CG. 
\newblock{A SIR model assumption for the spread of COVID-19 in different communities}.
\newblock Chaos, Solitons \& Fractals. 2020;139:110057.


\bibitem{DL_time_review}
Fawaz HI, Forestier G, Weber J, Idoumghar L, Muller PA.
\newblock{Deep learning for time series classification: a review}
\newblock Data Mining and Knowledge Discovery. 2019;33(4):917--963.


\bibitem{undersampling_loss_combine}
Kotsiantis S, Kanellopoulos D, Pintelas P, et al. 
\newblock{Handling imbalanced datasets: A review}
\newblock GESTS International Transactions on Computer Science and Engineering. 2006;30(1):25--36.


\bibitem{butterworth}
Selesnick IW, Burrus CS.
\newblock{Generalized digital Butterworth filter design}.
\newblock IEEE Transactions on signal processing. 1998;46(6):1688--1694.


\bibitem{normalisation}
Kotsiantis SB, Kanellopoulos D.
\newblock{Data preprocessing for supervised leaning}
\newblock International journal of computer science. 2006;1(2):1--7.


\bibitem{LSTM}
Gers FA, Schmidhuber J, Cummins F.
\newblock{Learning to forget: Continual prediction with LSTM}.
\newblock 1999.


\bibitem{cost_vs_undersampling}
Weiss GM, McCarthy K, Zabar B.
\newblock{Cost-sensitive learning vs. sampling: Which is best for handling unbalanced classes with unequal error costs}
\newblock 2007;7(35-41):24.

\bibitem{adam}
Kingma DP, Ba J.
\newblock{Adam: A method for stochastic optimization}
\newblock arXiv preprint arXiv:14126980. 2014.
\end{thebibliography}
\end{document}